\documentclass[%
aps,
prd,
reprint,
superscriptaddress,
nofootinbib,
amsmath,
amssymb]{revtex4-1}

\usepackage{indentfirst}  
\usepackage{bm}    
\usepackage{graphicx}  
\usepackage{latexsym}
\usepackage{graphicx}
\usepackage{psfrag}
\usepackage{epsfig}
\usepackage{amsmath}
\usepackage{array}
\usepackage{setspace}
\usepackage{amssymb}
\usepackage{float}
\usepackage{multirow}
\usepackage{slashbox}
\usepackage{color}
\usepackage[config=altsf]{subfig}
\usepackage[titles,subfigure]{tocloft}
\begin{document}

\title{Electromagnetic counterparts of  high-frequency gravitational waves having additional polarization states:  distinguishing and probing tensor-mode, vector-mode and scalar-mode gravitons}   




\author{Fang-Yu Li}
\email[]{fangyuli@cqu.edu.cn}
\affiliation{Physics Department, Chongqing University, Chongqing 401331, China.}

\author{ Hao Wen}
\email[]{wenhao@cqu.edu.cn}
\affiliation{Physics Department, Chongqing University, Chongqing 401331, China.}

\author{Zhen-Yun Fang}
\affiliation{Physics Department, Chongqing University, Chongqing 401331, China.}

\author{Di Li}
\affiliation{National Astronomical Observatories, Chinese Academy of Science, A20 Datun Road, Chaoyang District, Beijing 100012, China}

\author{Tong-Jie Zhang}
\affiliation{Department of Astronomy, Beijing Normal University, Beijing 100875, China}

%
%
%
%
%


\date{\today}

\begin{abstract}
\indent Gravitational waves (GWs)  from extra dimensions, very early   universe, and some high-energy astrophysical process,  might have, at most six polarization states in our four-dimensional spacetime:  $\oplus$-type and $\otimes$-type polarizations 
(tensor-mode gravitons), $x$-type, $y$-type polarizations (vector-mode gravitons), and $b$-type, $l$-type polarizations (scalar-mode gravitons). The peak regions or partial peak regions (of the amplitudes or energy densities) of some of such GWs are just distributed in GHz or higher frequency band, which would be  optimal frequency band for the electromagnetic (EM) response.  In this paper we investigate the EM response to such high-frequency GWs (HFGWs) having additional polarizations.  For the first time we address: 
(1) the concrete forms of analytic solutions for the perturbative EM fields caused by the HFGWs having all  six possible polarizations  in the background stable EM fields;
(2) the perturbative EM signals of the HFGWs with additional polarizations in the three-dimensional synchro-resonance system (3DSR system) and in the galactic-extragalactic background EM fields. 
These perturbative EM fields are actually the EM counterparts of the HFGWs, and such results provide a novel way to simultaneously distinguish and display the all possible six polarizations states of the HFGWs. It is also shown that:
(i) In the EM response, the pure $\otimes$-type, pure $x$-type and pure $y$-type polarizations of the HFGWs can independently generate the perturbative photon fluxes (PPFs, i.e., the signal photon fluxes), while the $\oplus$-type, $b$-type and $l$-type polarizations produce the PPFs in different combination states of them. (ii) All such six polarization states of the HFGWs have the separability and detectability. (iii) In the EM response to the HFGWs from   the extra-dimensions, distinguishing and displaying the different polarization states of them would be quite possible due to their very high frequencies, large energy densities and special properties of spectrum. The observation and separation of the polarization states of the primordial HFGWs and of the HFGWs in some high-energy astrophysical processes will face to big challenge, but this is still possible. (iv) Detection frequency band ($\nu\sim 10^{8} Hz$ to $10^{12} Hz$ or higher) of the PPFs by the 3DSR system and the observation frequency range ($\nu\sim 7\times10^{7} Hz$ to $3\times10^{9} Hz$) of the PPFs by the FAST (Five-hundred-meter Aperture Spherical Telescope, China), have a certain overlapping property, and thus their coincidence experiments for observations will have high complementarity. \\

\textbf{{Keywords}}: gravitational waves, high-frequency gravitational waves,electromagnetic counterparts, additional polarizations

\begin{description}
	\item[PACS numbers]
	04.30.Nk, 04.50.-h, 04.25.Nx, 04.80.Cc
\end{description}
\end{abstract}

\maketitle

\section{Introduction}
\label{Intro}

\indent Recently, the LIGO Scientific Collaboration and the Virgo Collaboration reported multiple gravitational wave (GW) evidences,e,g. (GW150914, GW151226, GW170104, GW170608,  GW170814, GW170817)\cite{PhysRevLett.116.061102,secondLIGOGW,PhysRevLett.118.221101,GW170608,GW170814,bns} and a candidate (LVT151012)\cite{PRX041015}etc.  These GW events and candidate are mainly produced by binary black hole mergers (frequencies are distributed around 30Hz to 450Hz, and dimensionless amplitudes in region of the Earth are $h\sim10^{-21}$ to $h\sim10^{-22}$), and one event of them is the detection evidence of GWs produced by the binary neutron star merger\cite{bns}. Obviously, they are the most important achievements since the observation of the gravitational radiation from PSR1913+16 for the GW project. Because the binary neutron star merger was accompanied by gamma-ray radiation almost at the same time, it is actually an electromagnetic (EM) counterpart of the GWs. In fact, these EM counterparts may exist widely in the universe, and thus they will provide a new and effective tool for observation and detection of the GWs. Obviously, the above-mentioned results provide the direct evidence for detection of the GWs expected by the GR. Moreover, these achievements have also following important scientific significance:\\
\indent1. They provide effective tools for exploration of the merger mechanism of compact objects, such as the black holes and neutron stars.\\
\indent2. Amplitude magnitude of $h\sim10^{-21}$ to $h\sim10^{-22}$  for such GWs shows the rationality of weak field linear approximation and the perturbation theory of gravity. Thus it further increases the possibility and hope of searching gravitons of spin-2 in the quantization process of gravity.\\
\indent3. Successful detection of the GWs by the ground-based GW detectors such as LIGO and Advanced Virgo indicates the validity of tetrad coordinates, i.e., observable quantities should be the projections of the physical quantities of the GWs as a tensor on tetrads of the observer's world-line.\\
\indent4. The information (including related energy-momentum) carried by the GWs from the wave sources of the binary compact objects also provides a strong evidence of the positive definite property of the energy-momentum tensor for the GW fields themselves.\\
\indent5. Observation of the GWs emitted by the binary neutron star merger and related EM counterpart (the gamma-ray radiation) gives a strict limit to the propagating velocity of the GWs in the intermediate frequency band ($\nu_g\sim$ 1 Hz to 1000 Hz). Also, it provides effective constraint for the geometry of extra dimensions.\\
\indent On the other hand, the results obtained by LIGO and Virgo should not mean the end of seeking the GWs. On the contrary, they are just the beginning of the GW astronomy. This is because of the following reasons:\\
\indent 1. With the continuous improvement of the sensitivities of LIGO, Virgo, GEO, KAGRA, AIGO etc., searchable sky area and detectable space scale will be further expanded. Therefore, it is very possible that more GW evidences will be detected and found during the next decade.\\
\indent 2. The GWs recently detected by LIGO and Virgo Collaboration are located in an interesting but special intermediate frequency range ($\nu\sim 30Hz$ to $\nu\sim 450Hz$ ), and their durations of signals in the detectors were very short. Thus, observation and detection to the continuous GWs, other kinds of GWs and the GWs in other frequency bands will be urgent affairs.\\
\indent 3. Except for the GWs predicted by General Relativity (GR), series of modified gravity theories and the gravity theories beyond GR also expect the GWs. These gravity theories include general metric theory of gravity\cite{PhysRevLett.30.884,Will}, Brans-Dicke theory\cite{PhysRev.124.925,Fujii}, scalar-tensor theories of gravity
\cite{arxiv1106.3312,PhysRevD.89.103008} and vector-tensor theories of gravity\cite{PLB2016}, f(R) gravity theory\cite{Capozziello2008}, and so on\cite{DVALI2000208, Seahra.PRL.2005, cqgF33,Andriot2017}. An important difference to the GR is that the GWs in some of such gravity theories, might have additional polarization states, which can be at most six polarization states   in our 3+1 dimension spacetime, while the GWs in the GR have only two polarization states, i.e., $\oplus$-type and $\otimes$-type polarization states. Thus, further theoretical study and experimental observation of the GWs, will provide important criterions for the polarization states,   the propagating speed, the waveforms, and other possible novel properties of the GWs.\\

\indent For the intermediate frequency GWs ($\nu\sim 1 Hz$ to $\nu\sim 10^{3} Hz$), Nishizawa et al investigated effective methods and schemes to detect and separate the different polarization states of such GWs. In fact, such schemes are based on the correlation analysis of multiple ground-based GW detectors\cite{PhysRevD.79.082002,PhysRevD.88.064005}. For the detection of low-frequency GWs ($\nu\sim 1 Hz$ to $\nu\sim 0.1 Hz$), related scheme is based on the configuration of space-based GW detectors\cite{PhysRevD.81.104043}. These schemes would be promising for displaying and separating the all polarization states of the GWs. Obviously, all of the above schemes are based on the tidal action caused by these polarization states of the GWs. Moreover, L Visinalli et al studied the scheme probing extra dimensions by the intermediate frequency GW signals and related EM counterparts from the mergers of the binary neutron stars \cite{arXiv1711.06628}. Such scheme and direct detection of the additional polarizations of the HFGWs from the extra dimensions, will be complementary in the frequency band and in the principles.\\
\indent On the other hand, almost all mainstream inflationary theories and universe models predicted the primordial (relic) GWs, and the spectrum of these relic GWs would distribute in a very wide frequency region, which may be from extreme-low frequency range ( $\nu\sim 10^{-16} Hz$ to $\nu\sim 10^{-17} Hz$) to high-frequency band ( $\nu\sim 10^{8} Hz$ to $\nu\sim 10^{10} Hz$ or higher)\cite{0504018,084022,pr373,sciam290,prd123511,cqg045004,Giovannini2014,Ito.anisotropic.2016}. Especially, the peak region or partial peak region of the energy densities of the relic HFGWs predicted by the pre-big-bang models\cite{pr373,sciam290}, the quintessential inflationary models\cite{prd123511,cqg045004,Giovannini2014}, and the short-term anisotropic inflation model\cite{Ito.anisotropic.2016}, are just distributed in the typical microwave band ( $\nu\sim 10^{8} Hz$ to $\nu\sim 10^{10} Hz$), in which corresponding amplitudes might reach up to $h\sim10^{-26}$ to $h\sim10^{-30}$. Moreover, the frequency of the HFGWs (KK-gravitons) expected by the braneworld scenarios\cite{cqgF33,Andriot2017} in extra-dimensions and the HFGWs predicated by interaction of astrophysical plasma with intense electromagnetic waves (EMWs) \cite{prd044017}, have been extended to  $\sim 10^{9} Hz$ to  $\sim 10^{12} Hz$ or higher (related amplitudes of these HFGWs would be expected to be $h\sim10^{-21}$ to $\sim10^{-27}$\cite{cqgF33,Andriot2017,prd044017}), and some works predicted HFGWs in very high-frequency band even over $10^{19}Hz$ from coherent oscillation of
electron-positron pairs and fields\cite{HanPRD89.024008} or from the magnetars\cite{WenMagnetar2017}. Another important possible source of HFGWs is produced during the preheating in the early universe and it would have frequency over GHz\cite{liujing2018,Easther}. Mergers\cite{LiXinPrimBH2017} and evaporation\cite{Fujita2014} of primordial black holes are also possible HFGW sources, and their frequency band can be $\sim10^9Hz$ to $10^{13}Hz$, in which, related amplitudes might be $h\sim10^{-31}$ to $h\sim10^{-36}$, respectively. In fact, according to the GR, and even modified gravity theories and the gravity theories beyond the GR, any energy-momentum tensor in the high-frequency oscillating states with deviation from the spherical symmetry or cylindrical symmetry, would be possible and potential HFGW sources. Thus such mechanisms and process would be very common in the high-energy astrophysics and cosmology, and these HFGWs would contain abundant astrophysical and cosmological information.\\
\indent However, the frequencies of these HFGWs are far beyond the detection or observation range of the intermediate-frequency GWs ($\nu_g\sim 1 Hz$ to $\sim 10^3 Hz$, by ground-based GW detectors), the low-frequency GWs ( $\nu_g\sim 1 Hz$ to $\sim 10^{-7} Hz$, by space GW detectors) and the extreme-low frequency GWs ( $\nu_g\sim 10^{-16} Hz$ to $\sim 10^{-17} Hz$, by B-mode polarization in the CMB). Therefore, detection and observation of such HFGWs need new principle and scheme, including separation of the additional polarization states of the HFGWs. So far, the separation and distinguishing of the additional polarization states of the HFGWs in the microwave frequency band by the electromagnetic (EM) response, almost have not been reported in the past.\\
\indent In this paper, based on the electrodynamics and the quantum electronics in curved spacetime, we investigate  a novel way to distinguish and display the all possible six polarization states of the HFGWs. 
We will address the concrete forms of analytic solutions for the perturbative EM fields caused by the HFGWs having all  six possible polarizations in the background stable EM fields, and study the perturbative EM signals (actually they are also the EM counterparts) of the HFGWs with additional polarizations in the 3DSR system (in laboratory scale) and in galactic-extragalactic background EM fields.
Our attention will focus mainly on the EM response to the HFGWs from the extra-dimensions, from very early stage of the universe and from some high-energy astrophysical process, especially the HFGWs (KK-gravitons) from the braneworld\cite{cqgF33,Andriot2017}, the relic HFGWs from the short-term anisotropic inflation\cite{Ito.anisotropic.2016}, from the pre-big-bang\cite{pr373,sciam290}, from the quintessential inflation\cite{prd123511,cqg045004,Giovannini2014}, and the HFGWs from the astrophysical plasma oscillation\cite{prd044017}, etc. This is because they involve following important scientific issues and problems: the extra-dimensions of space and the brane universes,  the very early universe and inflationary epoch, the start point of time or the information from the pre-big-bang, the essence and candidates of dark energy, and the interaction mechanism of the astrophysical plasma with intense EM radiations.\\
\indent The plan of this paper is organized as follows. 
In sec. \ref{addpolrztstates}, we shall show the general form of the HFGWs having six polarization states.
In sec. \ref{perturbativeEMWs}, we address the analytic solutions for the perturbative EM fields caused by HFGWs having six polarization states in the background stable EM fields, and the separation and displaying of such six polarizations. 
In sec. \ref{3DSR}, we study the EM signals caused by the six polarizations of HFGWs in the 3DSR system (in laboratory scale), and their separation and displaying effects. 
In sec. \ref{numerical}, numerical estimations of the perturbative photon fluxes in the 3DSR system and in the galactic-extragalactic magnetic fields, are given. Our brief conclusion is summarized in \mbox{sec. \ref{conclusion}}.

\section{High-frequency gravitational wave (HFGWs) having additional polarization states.}
\label{addpolrztstates}
\indent 1. In general, the ``monochromatic components'' of the GWs having six polarization states and propagating along the z-direction can be written as
\begin{eqnarray}\label{eq01}
	&~&h_{\mu\nu} =\left( {{\begin{array}{*{20}c}
				0 &  0   & 0 &0\\
				0 & A_{\oplus}+A_b&A_{\otimes}  &A_x\\
				0 & A_{\otimes}  &  -A_{\oplus}+A_b& A_y\\
				0 &  A_x   & A_y& \sqrt{2}A_l \nonumber\\
	\end{array} }} \right)\exp[i(k_gz-\omega_gt)],\nonumber\\
\end{eqnarray}
where $\oplus$, $\otimes$, x, y, b and $l$ represent $\oplus$-type, $\otimes$-type polarizations (tensor-mode gravitons), x-type, y-type polarizations (vector-mode gravitons), b-type and $l$-type polarizations (scalar-mode gravitons), respectively. For the coherent and non-stochastic GWs, such as the GWs from extra dimensions (e.g., the \mbox{K-K} GWs from braneworld\cite{cqgF33,Andriot2017}), the  $A_{\oplus}$,  $A_{\otimes}$, $A_{x}$, $A_{y}$, $A_{b}$ and $A_{l}$ are constant values of amplitudes of the GWs in the laboratory frame of reference. For the relic GWs, the $A_{\oplus}$, $A_{\otimes}$, $A_{x}$, $A_{y}$, $A_{b}$ and $A_{l}$ are  stochastic values of the amplitudes of the relic GWs in the laboratory frame of reference, which contain the cosmology scale factor; the $k_g$ and $\omega_g$ are wave number and angular frequency of the GWs, respectively. \\
\indent Eq.(\ref{eq01}) can be conveniently represented as following matrix form\cite{PhysRevD.79.082002}
\begin{eqnarray}\label{eq02}
	&~&e_{ij}^{\oplus} =\left( {{\begin{array}{*{20}c}
				1 &  0   & 0 \\
				0 &  -1   & 0 \\
				0 &  0   & 0 \\
				
	\end{array} }} \right),~~~~~~e_{ij}^{\otimes} =\left( {{\begin{array}{*{20}c}
				0 &  1   & 0 \\
				1 &  ~0   & ~0 \\
				0 &  0   & 0 \\
	\end{array} }} \right) 
	\nonumber\\~	
	&~&e_{ij}^{x} =\left( {{\begin{array}{*{20}c}
				0 &  0   & 1 \\
				0 &  ~0   & ~0 \\
				1 &  0   & 0 \\
				
	\end{array} }} \right),~~~~~~e_{ij}^{y} =\left( {{\begin{array}{*{20}c}
				0 &  0   & 0 \\
				0 &  ~0   & ~1 \\
				0 &  1   & 0 \\
				
	\end{array} }} \right) \nonumber\\
	\nonumber\\~	
	&~&e_{ij}^{b} =\left( {{\begin{array}{*{20}c}
				1 &  0   & 0 \\
				0 &  ~1   & ~0 \\
				0 &  0   & 0 \\
				
	\end{array} }} \right),~~e_{ij}^{l} =\sqrt{2}\left( {{\begin{array}{*{20}c}
				0 &  0   & 0 \\
				0 &  ~0   & ~0 \\
				0 &  0   & 1 
	\end{array} }} \right) 
\end{eqnarray}\\
\indent 2. Phase modification issues in the EM response to the HFGWs.\\
\indent Here, the above expressions are based on following assumption, i.e., the all components of the GWs have the same propagating velocity (the speed of light). In this case, dose it mean that the above gravitons must be massless (i.e., must be gravitons predicted by the GR)?  The answer to this question is that it depends on different cases. In fact, according to a series of modified and extended gravitational theories, the gravitons might have very small but non-vanishing masses, and the propagating velocities of such gravitons (e.g., massive gravitons) will be different to the speed of light. However, even if there is such difference,  the modifications to the propagating velocity of the GWs (gravitons) will be extremely small values, and the phase difference between the GWs and the perturbative EM waves (photons)  produced by the EM response to the GWs can often be neglected in the propagating distance discussed in this paper, especially, for the HFGWs having higher frequencies.\\
\indent According to the modified and extended gravity theories \cite{PhysRevD.88.064005,PhysRevD.85.043005}, the group velocity and the phase velocity of the gravitons would be:
\begin{eqnarray}\label{eq05}
	v_{_G}=c\sqrt{1-\frac{m_g^2c^4}{\hbar^2\omega_g^2}},
\end{eqnarray}
\begin{eqnarray}\label{eq06}
	\text{and,}~~~~	v_{_P}=c \left(\sqrt{1-\frac{m_g^2c^4}{\hbar^2\omega_g^2}}\right)^{-1},
\end{eqnarray}
where $m_g$, $\omega_g$ are the mass of graviton and the angular frequency, respectively.\\
\indent In the EM response to the HFGWs, it can be shown that the phase modification caused by the difference between the phase velocity $v_{_P}$ of the HFGWs (gravitons) and the phase velocity $c$ of the electromagnetic wave  (EMW, i.e., photon fluxes) can often be neglected, even if in the typical astrophysical distance. According to Eq.(\ref{eq06}) and propagating factor $exp[i(k_gz-\omega_gt)]$ in Eq.(\ref{eq01}), the final accumulation phase modification $\bigtriangleup\Phi_f$ caused by the HFGWs (gravitons) and the EM signals (perturbative photon fluxes, see below) can be given by:
\begin{eqnarray}\label{eq07}
	\bigtriangleup\Phi_f=(k_g-k_{_P})\bigtriangleup z=\omega_g(\frac{\bigtriangleup z}{c}-\frac{\bigtriangleup z}{v_{_P}})\nonumber\\
	=\frac{\omega_g \bigtriangleup z}{c}\left(1-\sqrt{1-\frac{m_g^2c^4}{\hbar^2\omega_g^2}}\right),
\end{eqnarray}
where $\bigtriangleup z$ is the distance of the GW source to the Earth. If $m_g^2c^4\ll\hbar^2\omega_g^2$ (for usual estimation of the possible values of the masses for gravitons, such condition can always be valid, especially the gravitons having high-frequency, i.e., the HFGWs). Then Eq.(\ref{eq07}) can be reduced to
\begin{eqnarray}\label{eq08}
	\bigtriangleup\Phi_f \approx\frac{\omega_g \bigtriangleup z}{2c}\cdot\frac{m_g^2c^4}{\hbar^2\omega_g^2}\propto \frac{m_g^2}{\omega_g},
\end{eqnarray}
\indent This means that smaller mass $m_g$ and higher frequency $\omega_g$ will have a better coherence resonance effect between the HFGWs (gravitons) and the perturbative EM fields (signal photon fluxes).\\
\indent Table \ref{table01} shows that in most cases such phase modification would be much less than $2\pi\times10^{-3}$. Notice that even if $\Delta z\sim10^{21}m$, which is equivalent to the diameter of the Galaxy, the final phase difference will still be less than or much less than $\sim 10^{-3}\times2\pi$. Thus the assumption of $v_g=c$ in Eq.(\ref{eq01})  is reasonable and valid in the typical distance discussed in this paper. Obviously, the above result is exactly valid for the GWs in the framework of GR.\\
\begin{table*}[!htbp]
	\caption{\label{table01}%
		Final phase modification in the typical propagating distance $\Delta z$ (the distance of the HFGW sources to the Earth) discussed in this paper due to the possible graviton masses. For the possible HFGW sources in the braneworld\cite{cqgF33}, such distance is estimated as $\Delta z\sim 10^{19}m$ and $\sim 10^{21}$m.  $\Delta\phi_f$ is the final phase difference.}
	\begin{tabular}{cccccc}
		\hline
		&  & \multicolumn{2}{c}{  $\Delta\phi_f$,   $\Delta z\sim 10^{19}m$} & \multicolumn{2}{c}{ $\Delta\phi_f$, $\Delta z\sim 10^{21}m$}  \\
		\cline{3-6}
		Method to predict the mass  & Upper limit &&&&\\
		limit of massive graviton&of mass(kg) & for $\nu\sim10^{12}Hz$& for $\nu\sim10^{9}Hz$&for $\nu\sim10^{12}Hz$ & for $\nu\sim10^{9}Hz$ \\
		\hline
		Far field constraints\cite{PhysRevD.68.024012}& $10^{-69}$ & $3.4\times10^{-38}\cdot2\pi$ & $3.4\times10^{-35}\cdot2\pi$& $3.4\times10^{-36}\cdot2\pi$ & $3.4\times10^{-33}\cdot2\pi$ \\
		Near field constraints\cite{Gruzinov2005311}& $10^{-67}$ & $3.4\times10^{-34}\cdot2\pi$ & $3.4\times10^{-31}\cdot2\pi$&$3.4\times10^{-32}\cdot2\pi$ & $3.4\times10^{-29}\cdot2\pi$ \\
		Pulsar timing arrays\cite{0004-637X-722-2-1589} & $3\times10^{-59}$ & $3.0\times10^{-17}\cdot2\pi$  & $3.0\times10^{-14}\cdot2\pi$ &$3.0\times10^{-15}\cdot2\pi$  & $3.0\times10^{-12}\cdot2\pi$\\
		Pulsar timing\cite{PhysRevD.78.044018} & $2\times10^{-59}$ & $1.4\times10^{-17}\cdot2\pi$  & $1.4\times10^{-14}\cdot2\pi$ & $1.4\times10^{-15}\cdot2\pi$  & $1.4\times10^{-12}\cdot2\pi$\\
		Binary pulsars\cite{PhysRevD.65.044022}& $7.6\times10^{-56}$ & $2.0\times10^{-10}\cdot2\pi$ & $2.0\times10^{-7}\cdot2\pi$&$2.0\times10^{-8}\cdot2\pi$ & $2.0\times10^{-5}\cdot2\pi$ \\
		\hline
	\end{tabular}
\end{table*}
\indent 3. The metric components of the HFGWs having six polarization states.\\
\indent It is well known that for the weak GWs, the metric can often be expressed as a small perturbation to the background spacetime $\eta_{\mu\nu}$, i.e., 
\begin{eqnarray}\label{eq09}
	g_{\mu\nu}=\eta_{\mu\nu}+h_{\mu\nu}
\end{eqnarray}
From Eqs.(\ref{eq01}) and (\ref{eq02}), the covariant   components of the metric tensor for the HFGWs can be given by
\begin{eqnarray}\label{eq10}
	&&	g_{00}=-1, g_{01}=g_{02}=g_{03}=g_{10}=g_{20}=g_{30}=0, \nonumber\\
	&&	g_{11}=1+h_{11}=1+h_{\oplus}+h_{b},~g_{22}=1+h_{22}=1-h_{\oplus}+h_{b}\nonumber\\
	&&	g_{12}=g_{21}=h_{12}=h_{21}=h_{\otimes},~g_{13}=g_{31}=h_{13}=h_{31}=h_x,\nonumber\\
	&&	g_{23}=g_{32}=h_{23}=h_{32}=h_{y},~g_{33}=1+h_{33}=1+h_{l}, 
\end{eqnarray}
where 
\begin{eqnarray}\label{eq12}
	\text{det}|g_{\mu\nu}|=1+h_{11}+h_{22}+h_{33}+O(h^2)\approx 1+2h_b+h_l,  ~~~~~ 
\end{eqnarray}
In the above expressions, we neglected the second-order infinite small quantity $h^2$ of the perturbation $h_{ij}$.\\
\indent Obviously, the HFGWs expressed by Eqs.(\ref{eq10}) and (\ref{eq12}) do not satisfy the transverse and traceless gauge condition (TT-gauge condition). Different polarization states in Eq. (\ref{eq01})   corresponds to different kinds of gravitons, where the $\oplus$-type and $\otimes$-type polarization states represent gravitons of spin-2; the x-type and y-type polarization states indicate gravitons of spin-1, and the b-type and $l$-type polarization states denote gravitons of spin-0, respectively. \\
\indent In fact, the EM perturbation effects produced by the HFGWs and their detections had been discussed. These EM detection systems include constructed and proposed ones like the two coupled spherical cavities\cite{Ballantini_CQG3505_2003}, high-frequency phonon trapping acoustic cavities\cite{PhysRevD.90.102005}, closed cylindrical superconducting cavity\cite{0306013} and coupling system between the planar superconducting open cavity and static magnetic field\cite{16}. However, in these previous researches, the HFGWs interacting with such EM systems are almost the GWs in the GR framework, i.e., they satisfy the TT-guage condition. So far,  it is not clear what a  concrete form (of the EM response to the HFGWs having additional polarization states) should be; it is also not clear whether such different polarization states of the HFGWs can be separated and displayed, and how to separate and display them. We shall show these issues in the expanded EM systems (see sect. \ref{3DSR}), which consist of not only static magnetic fields, but also static electric fields. Moreover, they also include the coupling system between the Gaussian-type microwave photon flux (Gaussian beam) and the  static high EM fields. Then, the EM perturbative effects produced by the different polarization states of the HFGWs will have different physical behaviors. Therefore, these different polarization states, in principle, can be separated and displayed.\\

\section{The perturbative effects of the HFGWs having six polarization states to the background stable EM fields}
\label{perturbativeEMWs}
\begin{figure}[!htbp]
	\centerline{\includegraphics[scale=0.6]{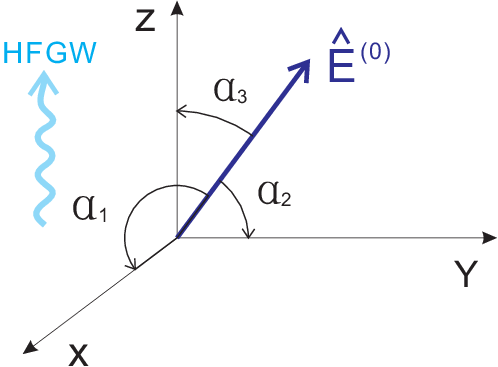}}
	\caption{\footnotesize{The background stable electric field $\hat{E}^{(0)}$ has an arbitrary direction, and $\alpha_1$, $\alpha_2$ and $\alpha_3$ are their directional factors. Here the HFGW propagates along the z-direction. }}
	\label{figure01}
\end{figure}

\begin{figure}[!htbp]
	\centerline{\includegraphics[scale=0.6]{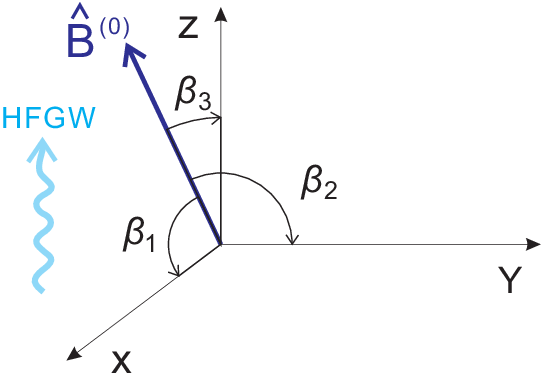}}
	\caption{\footnotesize{The background stable magnetic field $\hat{B}^{(0)}$ has also an arbitrary direction, and $\beta_1$, $\beta_2$ and $\beta_3$ are their directional factors. This magnetic field and the background electric field in Fig. \ref{figure01} are in the same region the HFGW passes through, and the HFGW propagates along the z-direction. }}
	\label{figure02}
\end{figure}

1. The perturbative solutions of the EM fields.\\
\indent In the EM response, the electric and magnetic fields can present at the same time, and they can point in arbitrary directions (see Figs.\ref{figure01} and \ref{figure02}).\\

\indent In Eqs. (\ref{eq13}), (\ref{eq21}) and Figs. \ref{figure01}, \ref{figure02}, the  ``$\wedge$'' stands the stable EM fields, and superscript ``0'' denotes the background EM fields. From Figs.  \ref{figure01}  and \ref{figure02}, the background stable EM fields can be written as following component forms:
\begin{eqnarray}\label{eq13}
	\hat E_x^{(0)}=\hat E^{(0)}\cos \alpha_1,~\hat E_y^{(0)}=\hat E^{(0)}\cos \alpha_2,~\hat E_z^{(0)}=\hat E^{(0)}\cos \alpha_3, \nonumber\\
	\hat B_x^{(0)}=\hat B^{(0)}\cos \beta_1,~\hat B_y^{(0)}=\hat B^{(0)}\cos \beta_2,~\hat B_z^{(0)}=\hat B^{(0)}\cos \beta_3,  ~~~~~
\end{eqnarray}
\begin{eqnarray}\label{eq14}
	\text{and,}~~~~~	\cos^2\alpha_1+\cos^2\alpha_2+\cos^2\alpha_3=1, \nonumber\\
	\cos^2\beta_1+\cos^2\beta_2+\cos^2\beta_3=1. 
\end{eqnarray}
According to Eqs. (\ref{eq13})  and  (\ref{eq14}), the covariant and the contra-variant components of the background EM field tensor can be given (we use MKS units):  $F_{01}^{(0)}=-F_{10}^{(0)}=\frac{1}{c}\hat E^{_{(0)}}_x$, $F_{02}^{(0)}=-F_{20}^{(0)}=\frac{1}{c}\hat E^{_{(0)}}_y$,
$F_{03}^{(0)}=-F_{30}^{(0)}=\frac{1}{c}\hat E^{_{(0)}}_z$,
$F_{12}^{(0)}=-F_{21}^{(0)}=-\hat B^{_{(0)}}_z$,
$F_{13}^{(0)}=-F_{31}^{(0)}=\hat B^{_{(0)}}_y$,
$F_{23}^{(0)}=-F_{32}^{(0)}=-\hat B^{_{(0)}}_x$.
\begin{eqnarray}
	\label{eq16}
	F^{\mu\nu(0)}=\eta^{\mu\alpha}\eta^{\nu\beta}F^{(0)}_{\alpha\beta}
\end{eqnarray}
Interaction of the HFGWs, Eq.(\ref{eq01}), with such background EM fields, Eqs. (\ref{eq13}) and (\ref{eq16}), will generate the EM perturbation, and the perturbative effects can be calculated by the electrodynamics equations in curved spacetime:
\begin{eqnarray}
	\label{eq17}
	&~& \frac{1}{\sqrt{-g}}\frac{\partial}{\partial x^{\nu}}[\sqrt{-g}g^{\mu\alpha}g^{\nu\beta}(F_{\alpha\beta}^{(0)}+\tilde F_{\alpha\beta}^{(1)})]=\mu_0J^{\mu}, \\
	\label{eq18}
	&~& \nabla_{\mu}F_{\nu\alpha}+\nabla_{\nu}F_{\alpha\mu}+\nabla_{\alpha}F_{\mu\nu}=0, \\
	\label{eq19}
	&~& \text{where,}~~\nabla_{\alpha} F_{\mu\nu}=F_{\mu\nu,\alpha}-\Gamma^{\sigma}_{\mu\alpha}F_{\sigma\nu}-\Gamma^{\sigma}_{\nu\alpha}F_{\mu\sigma},
\end{eqnarray}
$\Gamma^{\alpha}_{\mu\nu}$ is the second kind of Christoffel connection, and $J^{\mu}$ indicates the four-dimensional electric current density.  \\
\indent For the EM perturbation in the free space (vacuum), it has neither a real four-dimensional electric current nor other equivalent electric current, so $J^{\mu}=0$ in Eq.(\ref{eq17}). Moreover, for the EM perturbation generated by the weak HFGWs, we have $F_{\mu\nu}=\hat F_{\mu\nu}^{(0)}+\tilde F_{\mu\nu}^{(1)}$, where $\hat F_{\mu\nu}^{(0)}$ is the background EM field tensor, Eqs.(\ref{eq13}) and (\ref{eq16}), and $\tilde F_{\mu\nu}^{(1)}$ is the first-order perturbation to $F_{\mu\nu}^{(0)}$; the ``$\sim$'' represents time-dependent perturbative EM fields. Here we neglected the second-order and higher-order infinite small perturbations. In this case, we have:
\begin{eqnarray}
	\label{eq21}
	F_{\mu\nu}=  \hat F_{\mu\nu}^{(0)}+ \tilde F_{\mu\nu}^{(1)}+O(h^2),
\end{eqnarray}
\indent Introducing Eqs. (\ref{eq01}), (\ref{eq02}), (\ref{eq10} to \ref{eq13}) and (\ref{eq16})  into Eqs.(\ref{eq17}), (\ref{eq18}), we obtain following inhomogeneous hyperbolic-equations:
\begin{eqnarray}
	\label{eq22}
	\Box \tilde E_x^{(1)}&=&\frac{\partial^2\tilde E_x^{(1)}}{\partial z^2}-\frac{1}{c^2}\frac{\partial^2\tilde E_x^{(1)}}{\partial t^2}\nonumber\\
	&=&k_g^2\{(h_{\oplus}-\frac{1}{2}h_l)\hat E_x^{(0)} +  h_{\otimes} \hat E_y^{(0)} + h_{x} \hat E_z^{(0)}  \}\nonumber\\ 
	&+&ck_g^2\{ h_{\otimes} \hat B_x^{(0)} - (h_{\oplus}+\frac{1}{2}h_l)\hat B_y^{(0)} + h_{y} \hat B_z^{(0)}  \}, ~~~~~~~~
\end{eqnarray}
\begin{eqnarray}
	\label{eq23}
	\Box \tilde B_y^{(1)}&=&\frac{\partial^2\tilde B_y^{(1)}}{\partial z^2}-\frac{1}{c^2}\frac{\partial^2\tilde B_y^{(1)}}{\partial t^2}\nonumber\\
	&=&k_g^2/c\{(h_{\oplus}-\frac{1}{2}h_l)\hat E_x^{(0)} +  h_{\otimes} \hat E_y^{(0)} + h_{x} \hat E_z^{(0)}  \}\nonumber\\ 
	&+&k_g^2\{ h_{\otimes} \hat B_x^{(0)} - (h_{\oplus}+\frac{1}{2}h_l)\hat B_y^{(0)} + h_{y} \hat B_z^{(0)}  \}, ~~~~~~~~
\end{eqnarray}
\begin{eqnarray}
	\label{eq24}
	\Box \tilde E_y^{(1)}&=&\frac{\partial^2\tilde E_y^{(1)}}{\partial z^2}-\frac{1}{c^2}\frac{\partial^2\tilde E_y^{(1)}}{\partial t^2}\nonumber\\
	&=&k_g^2\{h_{x} \hat E_x^{(0)}-(h_{\oplus}+\frac{1}{2}h_l)\hat E_y^{(0)} +   h_{y} \hat E_z^{(0)}  \}\nonumber\\ 
	&-&ck_g^2\{ (h_b-h_{\oplus}+h_l)\hat B_x^{(0)}+h_{\otimes}\hat B_y^{(0)} + h_x\hat B_z^{(0)} \},  ~~~~~~~
\end{eqnarray}
\begin{eqnarray}
	\label{eq25}
	\Box \tilde B_x^{(1)}&=&\frac{\partial^2\tilde B_x^{(1)}}{\partial z^2}-\frac{1}{c^2}\frac{\partial^2\tilde B_x^{(1)}}{\partial t^2}\nonumber\\
	&=&-k_g^2/c\{h_{x} \hat E_x^{(0)}-(h_{\oplus}+\frac{1}{2}h_l)\hat E_y^{(0)} +   h_{y} \hat E_z^{(0)}  \}\nonumber\\ 
	&+&k_g^2\{ (h_b-h_{\oplus}+h_l)\hat B_x^{(0)}+h_{\otimes}\hat B_y^{(0)} + h_x\hat B_z^{(0)} \}, ~~~~~
\end{eqnarray}
\begin{eqnarray}
	\label{eq26}
	\text{and,}~~\frac{\partial \tilde E_z^{(1)}   }{\partial t  } &=& \frac{\partial h_x  }{\partial t  } \hat E_x^{(0)} + \frac{\partial h_y  }{\partial t  } \hat E_y^{(0)}-\frac{\partial    }{\partial  t } (h_b-\frac{1}{2}h_l)\hat E_z^{(0)}, \nonumber\\
	\frac{\partial \tilde E_z^{(1)}   }{\partial z  }  &=& \frac{\partial h_x  }{\partial z  } \hat E_x^{(0)} + \frac{\partial h_y  }{\partial z  } \hat E_y^{(0)}-\frac{\partial    }{\partial  z } (h_b-\frac{1}{2}h_l)\hat E_z^{(0)}, \nonumber\\
	\frac{\partial \tilde B_z^{(1)}   }{\partial t  } &=&\frac{\partial \tilde B_z^{(1)}   }{\partial z }=0,
\end{eqnarray}
where ``$\Box$'' indicates the d'Alembertian.\\
\indent From Eqs.(\ref{eq22}) to (\ref{eq26}), after lengthy calculations, we obtain the general solutions of these equations as follows:
\begin{eqnarray}
	\label{eq27}
	\tilde E^{(1)}_x=&-&\dfrac{i}{2}k_gz\{ (h_{\oplus}-\frac{1}{2}h_l)\hat E^{(0)}_x +  h_{\otimes}\hat E^{(0)}_y +  h_x\hat E^{(0)}_z \}\nonumber\\
	&-&\dfrac{i}{2}k_gcz\{ h_{\otimes}\hat B^{(0)}_x-(h_{\oplus}+\frac{1}{2}h_l)\hat B^{(0)}_y +  h_y\hat B^{(0)}_z \}\nonumber\\
	&+& C_1\exp[i(k_gz+\omega_gt)]
\end{eqnarray}
\begin{eqnarray}
	\label{eq28}
	\tilde B^{(1)}_y=&-&\dfrac{i}{2}\dfrac{k_gz}{c}\{ (h_{\oplus}-\dfrac{1}{2}h_l)\hat E^{(0)}_x + h_{\otimes}\hat E^{(0)}_y + h_x\hat E^{(0)}_z \}\nonumber\\
	&-&\dfrac{i}{2}k_gz\{ h_{\otimes}\hat B^{(0)}_x-(h_{\oplus}+\dfrac{1}{2}h_l)\hat B^{(0)}_y + h_y\hat B^{(0)}_z \}\nonumber\\
	&+& C_2\exp[i(k_gz+\omega_gt)]
\end{eqnarray}
\begin{eqnarray}
	\label{eq29}
	\tilde E^{(1)}_y=&-&\dfrac{i}{2}k_gz\{ h_x\hat E^{(0)}_x -(h_{\oplus}+\dfrac{1}{2}h_l)\hat E^{(0)}_y +  h_y\hat E^{(0)}_z \}\nonumber\\
	&+&\dfrac{i}{2}k_gcz\{ (h_b-h_{\oplus}+h_l) \hat B^{(0)}_x+h_{\otimes}\hat B^{(0)}_y + h_x\hat B^{(0)}_z \}\nonumber\\
	&+& C_3\exp[i(k_gz+\omega_gt)]
\end{eqnarray}
\begin{eqnarray}
	\label{eq30}
	\tilde B^{(1)}_x=&&\dfrac{i}{2}\dfrac{k_gz}{c}\{ h_x\hat E^{(0)}_x -(h_{\oplus}+\dfrac{1}{2}h_l)\hat E^{(0)}_y +  h_y\hat E^{(0)}_z \}\nonumber\\
	&-&\dfrac{i}{2}k_gz\{ (h_b-h_{\oplus}+h_l) \hat B^{(0)}_x+h_{\otimes}\hat B^{(0)}_y + h_x\hat B^{(0)}_z \}\nonumber\\
	&+& C_4\exp[i(k_gz+\omega_gt)]
\end{eqnarray}
\begin{eqnarray}
	\label{eq31}
	\text{and,}~~	\tilde E^{(1)}_z=h_x \hat E^{(0)}_x + h_y \hat E^{(0)}_y + (\dfrac{1}{2}h_l-h_b)\hat E^{(0)}_z,~~~
\end{eqnarray}
\begin{eqnarray}
	\label{eq32}
	\tilde B^{(1)}_z=0.
\end{eqnarray}
The Eqs. (\ref{eq27}) to (\ref{eq32}) show that:\\
\indent (1) If the background stable EM fields have all spatial components (i.e., the $x-$, $y-$ and $z-$components, here the HFGW propagates along the $z-$direction), then not only the EM response to the $\oplus$-type, $\otimes$-type polarization states in the GR framework can be produced, but also the EM response to the additional polarization states (the $x$-, $y$-, $b$- and $l$-type polarization states) beyond the GR can be generated.\\
\indent (2) The transverse polarization components ($\tilde E_x^{(1)}$, $\tilde B_y^{(1)}$, $\tilde E_y^{(1)}$, $\tilde B_x^{(1)}$) of the perturbative EM fields have the space accumulation effect (i.e., their strengths are proportional to the propagating distance $z$ of the HFGW in the background EM fields). This is because the HFGW (gravitons) and the perturbative EM waves (photons) have the same or almost the same propagating velocity (see section \ref{addpolrztstates}). Thus, they can generate an optimal space accumulation effect in the propagating direction, i.e., they and the results\cite{nuovo129,prd2915} in the GR from the Feynman perturbative techniques and the Einstein-Maxwell equations are self-consistent.
However, Eqs. (\ref{eq27}) to  (\ref{eq31}) include new EM counterparts generated by the EM response to HFGWs by the longitudinal background EM fields $\hat E^{(0)}_z$ and  $\hat B^{(0)}_z$ [see Eqs. (\ref{eq27}) to  (\ref{eq30})] and the longitudinal perturbative EM fields, Eq. (\ref{eq31}). 
Moreover, even considering correction to the propagating velocity for the HFGWs having additional polarization states [see Eqs.(\ref{eq05}) to (\ref{eq08}) and Table \ref{table01}, e.g., massive gravitons], such correction to the space accumulation effect can still be neglected even  in the typical astronomical scale. Therefore, the final phase difference caused by the phase velocity $v_{_P}$ does not cause any essential impact to the space accumulation effect (see Table \ref{table01}).\\
\indent  (3) Eqs.(\ref{eq27}) to (\ref{eq30}) also show that the perturbative EM fields propagating along the opposite direction of the HFGW [i.e., the perturbative EM fields containing the propagating factor ($k_gz+\omega_gt$)] do not have such space accumulation effect. Obviously, this is a self-consistent result to the EM perturbation, because there is  no accumulation effect of energy for such EM fields in the opposite propagating direction of the HFGW.\\
\indent (4) Unlike usual planar EM waves without longitudinal polarizations in free space, here the perturbative EM fields have the longitudinal polarizations, Eq. (\ref{eq31}), and the longitudinal component is only from the EM response of the background electric fields, $\hat E_x^{(0)}$, $\hat E_y^{(0)}$ and $\hat E_z^{(0)}$ and not of the background magnetic fields $\hat B_x^{(0)}$, $\hat B_y^{(0)}$ and $\hat B_z^{(0)}$. Moreover, since there is a phase difference of $\pi/2$ between the transverse perturbative EM fields [ 
$\tilde E_x^{(1)}$, $\tilde B_y^{(1)}$,  $\tilde E_y^{(1)}$ and $\tilde B_x^{(1)}$, Eqs.(\ref{eq27}) to (\ref{eq30})] and the longitudinal perturbative electric field  $\tilde E_z^{(1)}$, Eq.(\ref{eq31}), then the average values with respect to time of the transverse Poynting vector (i.e., the Poynting vector in $x$- and $y$-directions) are equal to zero [see Eq. (\ref{eq61})]. Clearly, such results ensure the total momentum conservation in the interaction of the HFGW with the background EM fields. However, such longitudinal perturbative electric field $\tilde E_z^{(1)}$ will play an important role for displaying and distinguishing the additional polarization states (the $x$-, $y$-, $b$- and $l$-polarizations) of the HFGWs in the 3DSR system (see below sect. \ref{3DSR} and Appendix A).\\

\begin{figure}[!htbp]
	\centerline{\includegraphics[scale=0.7]{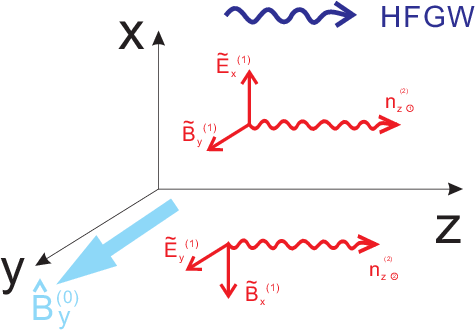}}
	\caption{\footnotesize{ When the HFGW, Eq.(\ref{eq01}), propagating in a transverse stable magnetic field $\tilde B^{(0)}_y$, the transverse perturbative EM fields, $ \tilde E^{(1)}_x$, $ \tilde B^{(1)}_y$, $ \tilde E^{(1)}_y$  and $ \tilde B^{(1)}_x$ can be generated, and they will produce the   PPFs in the propagating direction of the HFGW. The PPF $n_{z\tiny{\textcircled{1}}}^{(2)}$, Eq.(\ref{eq35}), generated by  $ \tilde E^{(1)}_x$ and $ \tilde B^{(1)}_y$ contain contribution from both the $\oplus$-type and the $l$-type polarizations, and the  PPF $n_{z\tiny{\textcircled{2}}}^{(2)}$, Eq.(\ref{eq36}), produced by  $ \tilde E^{(1)}_y$ and $ \tilde B^{(1)}_x$ contain only contribution from the $\otimes$-type polarization.  }}
	\label{figure03}
\end{figure}

\begin{figure}[!htbp]
	\centerline{\includegraphics[scale=0.7]{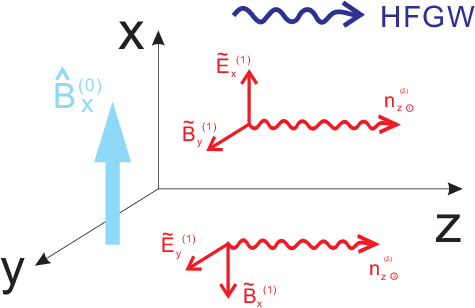}}
	\caption{\footnotesize{ When the HFGW  propagates in the transverse stable magnetic field $\tilde B^{(0)}_x$, the transverse perturbative EM fields, $ \tilde E^{(1)}_x$, $ \tilde B^{(1)}_y$, $ \tilde E^{(1)}_y$  and $ \tilde B^{(1)}_x$ can be generated, and they will produce the perturbative photon fluxes  $n_{z\tiny{\textcircled{1}}}^{(2)}$ and  $n_{z\tiny{\textcircled{2}}}^{(2)}$, Eqs.(\ref{eq39}) and (\ref{eq40}), in the propagating direction of the HFGW, where the $n_{z\tiny{\textcircled{1}}}^{(2)}$, Eq.(\ref{eq39}), contains only the contribution from the  $\otimes$-type polarization state, while  $n_{z\tiny{\textcircled{2}}}^{(2)}$, Eq.(\ref{eq40}), contains the contribution from the  $\oplus$-type, $b$-type and $l$-type polarization states of the HFGW. }}
	\label{figure04}
\end{figure}

\begin{figure}[!htbp]
	\centerline{\includegraphics[scale=0.7]{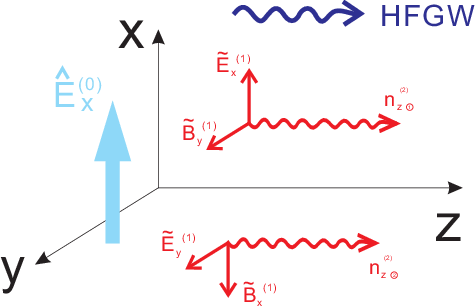}}
	\caption{\footnotesize{ When the HFGW  passes through the transverse stable electric field $\hat E^{(0)}_x$, the transverse perturbative EM fields, $ \tilde E^{(1)}_x$, $ \tilde B^{(1)}_y$, $ \tilde E^{(1)}_y$  and $ \tilde B^{(1)}_x$ can be generated, and they will produce the PPFs  $n_{z\tiny{\textcircled{1}}}^{(2)}$ and $n_{z\tiny{\textcircled{2}}}^{(2)}$ in the propagating direction of the HFGW.
			Here, the $n_{z\tiny{\textcircled{1}}}^{(2)}$, Eq.(\ref{eq45}), contains   the combined contribution from the  $\oplus$-type and the $l$-type polarization states (the tensor and the scalar mode gravitons) of the HFGW, while  $n_{z\tiny{\textcircled{2}}}^{(2)}$, Eq.(\ref{eq46}), contains only the contribution from the  pure $x$-type polarization state (the vector mode gravitons) of the HFGW. }}
	\label{figure05}
\end{figure}

\indent 2. Electromagnetic counterparts: the perturbative photon fluxes in the EM response.\\
\indent Interaction of the HFGWs with the EM fields will generate perturbative EM power fluxes (signal EM power fluxes). In order to conveniently represent the above EM signal in the background noise photon flux fluctuation (see Appendix B), we will express them in quantum language, i.e., the perturbative photon fluxes (the PPFs, or signal photon fluxes).
It is interesting to study following several cases:\\
\indent (1) The EM response to the HFGWs in transverse background stable magnetic field $\hat B_y^{(0)}$.\\
In this case,  $\hat E_x^{(0)}=\hat E_y^{(0)}=\hat E_z^{(0)}$$=\hat B_x^{(0)}=\hat B_z^{(0)}=0$ in Eqs.(\ref{eq27}) to (\ref{eq31}), i.e., only $\hat B_y^{(0)}$ has non-vanishing value. Then from  Eqs.(\ref{eq27}) to (\ref{eq30}), we obtain following results immediately,
\begin{eqnarray}
	\label{eq33}
	\tilde E^{(1)}_x&=&\dfrac{i}{2}k_gcz(h_{\oplus}+\frac{1}{2}h_l)\hat B_y^{(0)}\nonumber\\ 
	&=&\frac{i}{2} (A_{\oplus}+\frac{1}{2}A_l)\hat B_y^{(0)}k_gcz\exp[i(k_gz-\omega_gt)],\nonumber\\
	\tilde B^{(1)}_y&=&\dfrac{i}{2}k_gz(h_{\oplus}+\frac{\sqrt{2}}{2}h_l)\hat B_y^{(0)}\nonumber\\ 
	&=&\frac{i}{2} (A_{\oplus}+\frac{1}{2}A_l)\hat B_y^{(0)}k_gz\exp[i(k_gz-\omega_gt)] ,
\end{eqnarray} 
and 
\begin{eqnarray}
	\label{eq34}
	&&\tilde E^{(1)}_y=\dfrac{i}{2}k_gczh_{\otimes}\hat B_y^{(0)}=\frac{i}{2}A_{\otimes}\hat B_y^{(0)}k_gcz\exp[i(k_gz-\omega_gt)],\nonumber\\
	&&\tilde B^{(1)}_x=-\dfrac{i}{2}k_gzh_{\otimes}\hat B_y^{(0)}=-\frac{i}{2}A_{\otimes}\hat B_y^{(0)}k_gz\exp[i(k_gz-\omega_gt)],~~~~~~~
\end{eqnarray}
Here, as well as all discussions below, the perturbative EM fields propagating along the negative $z$-direction (i.e., the opposite propagating direction of the HFGWs) will be neglected, because they are very weak (they do not have the space accumulation effect) or absent\cite{nuovo129,prd2915}. Therefore, we can make the constants $C_1$, $C_2$, $C_3$ and $C_4$ equal to zero in Eqs. (\ref{eq27}) to (\ref{eq30}).\\
\indent The perturbative EM fields in Eqs. (\ref{eq33}), (\ref{eq34}) are similar to that in the refs. \cite{nuovo129,prd2915}. However, there is an important difference, i.e., the perturbative EM fields,  Eq.(\ref{eq33}),  contain not only the contribution from the $\oplus$-type polarization state of the HFGWs, but also the contribution from the longitudinal polarization (the $l$-type polarization) of the HFGWs. In other words, the contribution of the $\oplus$-type polarization state is always accompanied by the $l$-type polarization states. Moreover, they have the same symbol in the amplitudes of the perturbative EM fields,  Eq.(\ref{eq33}), i.e., this is a constructive coherence effect between the $\oplus$-type polarization and the $l$-type polarization states. Thus, the interaction of the HFGWs having additional polarization states with the background transverse magnetic fields $\hat B_y^{(0)}$, will consume more energy [see  Eq.(\ref{eq35})] than the HFGWs in the GR, i.e., the former will cause faster radiation damping than that of the latter for the HFGW sources in local regions.\\
\indent Unlike such effect, the perturbative EM fields,  Eq.(\ref{eq34}), produced by the $\otimes$-type polarization state of the HFGWs, does not contain the contribution from the $l$-type or other additional polarization states.\\
\indent From Eqs. (\ref{eq33}), (\ref{eq34}), the perturbative photon fluxes generated by the interaction of the HFGWs with the background magnetic field $\hat B_y^{(0)}$ can be given by (also see Fig. \ref{figure03}):
\begin{eqnarray}
	\label{eq35}
	n_{z\tiny{\textcircled{1}}}^{(2)}&=&\dfrac{1}{2\mu_0\hbar\omega_e}Re\langle \tilde E^{(1)*}_x \tilde B^{(1)}_y	\rangle     \nonumber\\
	 &=& \frac{1}{8\mu_0\hbar\omega_e}k_g^2z^2c[(A_{\oplus}+\frac{\sqrt{2}}{2}A_l)\hat B_y^{(0)}]^2, 
\end{eqnarray}
\begin{eqnarray}
	\label{eq36}
	n_{z\tiny{\textcircled{2}}}^{(2)}&=&\dfrac{1}{2\mu_0\hbar\omega_e} Re\langle \tilde E^{(1)*}_y \tilde B^{(1)}_x	\rangle          \nonumber\\
	 &=& \frac{1}{8\mu_0\hbar\omega_e}k_g^2z^2c(A_{\otimes} \hat B_y^{(0)})^2, ~~~~~~~~
\end{eqnarray}
where the * denotes complex conjugate; the angular brackets represent the average over time, and the superscript ``$2$'' represents second-order perturbation to the EM fields because they are proportional to the square of the HFGW amplitudes: $A_{\oplus}$, $A_{l}$ and $A_{\otimes}$. \\
\indent (2) The EM response to the HFGW in transverse background stable magnetic field $ \hat B^{(0)}_x$. Then $\hat E_x^{(0)}=\hat E_y^{(0)}=\hat E_z^{(0)}$$=\hat B_y^{(0)}=\hat B_z^{(0)}=0$  in Eqs. (\ref{eq27}) to (\ref{eq30}), i.e., only  $ \hat B^{(0)}_x$ is not equal to zero. In the same way, from Eqs. (\ref{eq27}) to (\ref{eq30}), corresponding perturbative photon fluxes are (see Fig. \ref{figure04}):
\begin{eqnarray}
	\label{eq39}
	n_{z\tiny{\textcircled{1}}}^{(2)}=\dfrac{Re\langle \tilde E^{(1)*}_x \tilde B^{(1)}_y	\rangle}{2\mu_0\hbar\omega_e} = \frac{k_g^2z^2c}{8\mu_0\hbar\omega_e}(A_{\otimes}\hat B_x^{(0)})^2,
\end{eqnarray}
\begin{eqnarray}
	\label{eq40}
	n_{z\tiny{\textcircled{2}}}^{(2)}=\dfrac{Re\langle \tilde E^{(1)*}_y \tilde B^{(1)}_x	\rangle}{2\mu_0\hbar\omega_e} = \frac{k_g^2z^2c}{8\mu_0\hbar\omega_e}[(A_b-A_{\oplus}+\sqrt{2}A_l)\hat B_x^{(0)}]^2, ~~~~~~
\end{eqnarray}
\indent Eq. (\ref{eq39}) is very similar to Eq. (\ref{eq36}), namely, the contribution of the $\otimes$-type polarization state is always independent of the additional polarization states. While, Eqs. (\ref{eq35}), (\ref{eq40}) show that the contribution of the $\oplus$-type polarization state is always accompanied by the $l$-type, or the $b$-type and $l$-type polarization states.\\
\indent So far, it is not clear yet the related ratio among the intensities of the  $\oplus$-type,  $b$-type and  $l$-type
polarization states. Obviously, in the following cases: (i) $A_{\oplus}\gg A_b, A_l$, (ii) $A_{b}\gg  A_{\oplus}, A_l$, or (iii) $A_{l}\gg A_{\oplus}, A_b$, then the effect of the $\oplus$-type polarization (effect in the GR),   the $b$-type or the $l$-type polarization (effect beyond the GR) can mainly be displayed, respectively.\\
\indent Notice that although $\hat B_y^{(0)}$ and $\hat B_x^{(0)}$ are all the transverse stable magnetic fields, their EM response to the HFGW have certain difference. The $n_{z\tiny{\textcircled{1}}}^{(1)}$, Eq.(\ref{eq35}), is a constructive coherence effect between the $\oplus$-type and the $l$-type polarization states (they have the same symbols), while the $n_{z\tiny{\textcircled{2}}}^{(1)}$, Eq.(\ref{eq40}), is a destructive coherence effect between the $\oplus$-type and the $b$-type, $l$-type polarization states (the $\oplus$-type  and the $b$-type, $l$-type polarization states have the opposite symbols). This is because the impacts of the $b$-type polarization to the $\oplus$-type polarization in the $xx$- and the $yy$-components of the HFGW metric $h_{\mu\nu}$ are different [see Eq.(\ref{eq01})]. The former is ``constructive superposition'' (where the $b$-type and the $\oplus$-type polarizations have the same symbols), and the latter is ``destructive superposition'' (where the $b$-type and the $\oplus$-type polarizations have the opposite symbols). Moreover, the $l$-type polarization only appears in the $zz$-component of the metric  $h_{\mu\nu}$. In this case, the EM response of $\hat B_x^{(0)}$ (the $yz$-component $F_{23}$ of the EM field tensor) and $\hat B_y^{(0)}$ (the $xz$-component $F_{13}$ of the EM field tensor) are non-symmetric. This is the physical origin of such difference.\\
\indent (3) The EM response to the HFGWs in the transverse background stable electric field $\hat E_x^{(0)}$. In this case,   $\hat E_y^{(0)}=\hat E_z^{(0)}=\hat B_x^{(0)}$$=\hat B_y^{(0)}=\hat B_z^{(0)}=0$ in Eqs. (\ref{eq27}) to (\ref{eq31}), i.e., only $\hat E_x^{(0)}$ has non-vanishing value. By using the same method, from Eqs. (\ref{eq27}) to (\ref{eq31}), we have (see Fig. \ref{figure05})
\begin{eqnarray}
	\label{eq45}
	n_{z\tiny{\textcircled{1}}}^{(2)}=\dfrac{Re\langle \tilde E^{(1)*}_x \tilde B^{(1)}_y	\rangle}{2\mu_0\hbar\omega_e} = \frac{k_g^2z^2}{8\mu_0c\hbar\omega_e}[(A_{\oplus}-\frac{\sqrt{2}}{2}A_l)\hat E_x^{(0)}]^2, ~~~~~~
\end{eqnarray}
\begin{eqnarray}
	\label{eq46}
	n_{z\tiny{\textcircled{2}}}^{(2)}=\dfrac{Re\langle \tilde E^{(1)*}_y \tilde B^{(1)}_x	\rangle}{2\mu_0\hbar\omega_e} = \frac{k_g^2z^2}{8\mu_0c\hbar\omega_e}(A_{x}\hat E_x^{(0)})^2,
\end{eqnarray}

\begin{figure}[!htbp]
	\centerline{\includegraphics[scale=0.7]{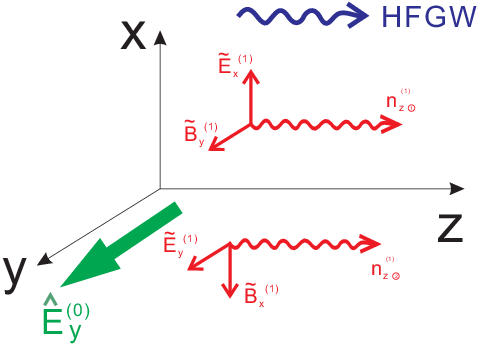}}
	\caption{\footnotesize{ When the HFGW  passes through the transverse stable electric field $\hat E^{(0)}_y$, the transverse perturbative EM fields, $ \tilde E^{(1)}_x$, $ \tilde B^{(1)}_y$, $ \tilde E^{(1)}_y$  and $ \tilde B^{(1)}_x$ can be generated, and they will produce the PPFs  $n_{z\tiny{\textcircled{1}}}^{(2)}$ and $n_{z\tiny{\textcircled{2}}}^{(2)}$ in the propagating direction of the HFGWs.
			Here, the $n_{z\tiny{\textcircled{1}}}^{(2)}$, Eq.(\ref{eq51}),    contains only the contribution from the the pure   $\otimes$-type   polarization states (the tensor  mode gravitons) of the HFGW, while  $n_{z\tiny{\textcircled{2}}}^{(2)}$, Eq.(\ref{eq52}), contains   the combined contribution from the    $\oplus$-type and $l$-type polarization state (the tensor and scalar mode gravitons) of the HFGW. }}
	\label{figure06}
\end{figure}

\begin{figure}[!htbp]
	\centerline{\includegraphics[scale=0.7]{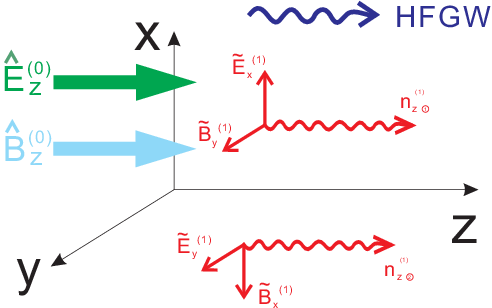}}
	\caption{\footnotesize{ When the HFGW  passes through the longitudinal stable EM fields $\hat B^{(0)}_z$ and $\hat E^{(0)}_z$ , the transverse perturbative EM fields, $ \tilde E^{(1)}_x$, $ \tilde B^{(1)}_y$, $ \tilde E^{(1)}_y$  and $ \tilde B^{(1)}_x$ can be generated, and they will produce the PPFs  $n_{z\tiny{\textcircled{1}}}^{(2)}$ and $n_{z\tiny{\textcircled{2}}}^{(2)}$,  Eqs.(\ref{eq57}) and (\ref{eq58}), in the propagating direction of the HFGW.}}
	\label{figure07}
\end{figure}

\begin{figure}[!htbp]
	\centerline{\includegraphics[scale=0.4]{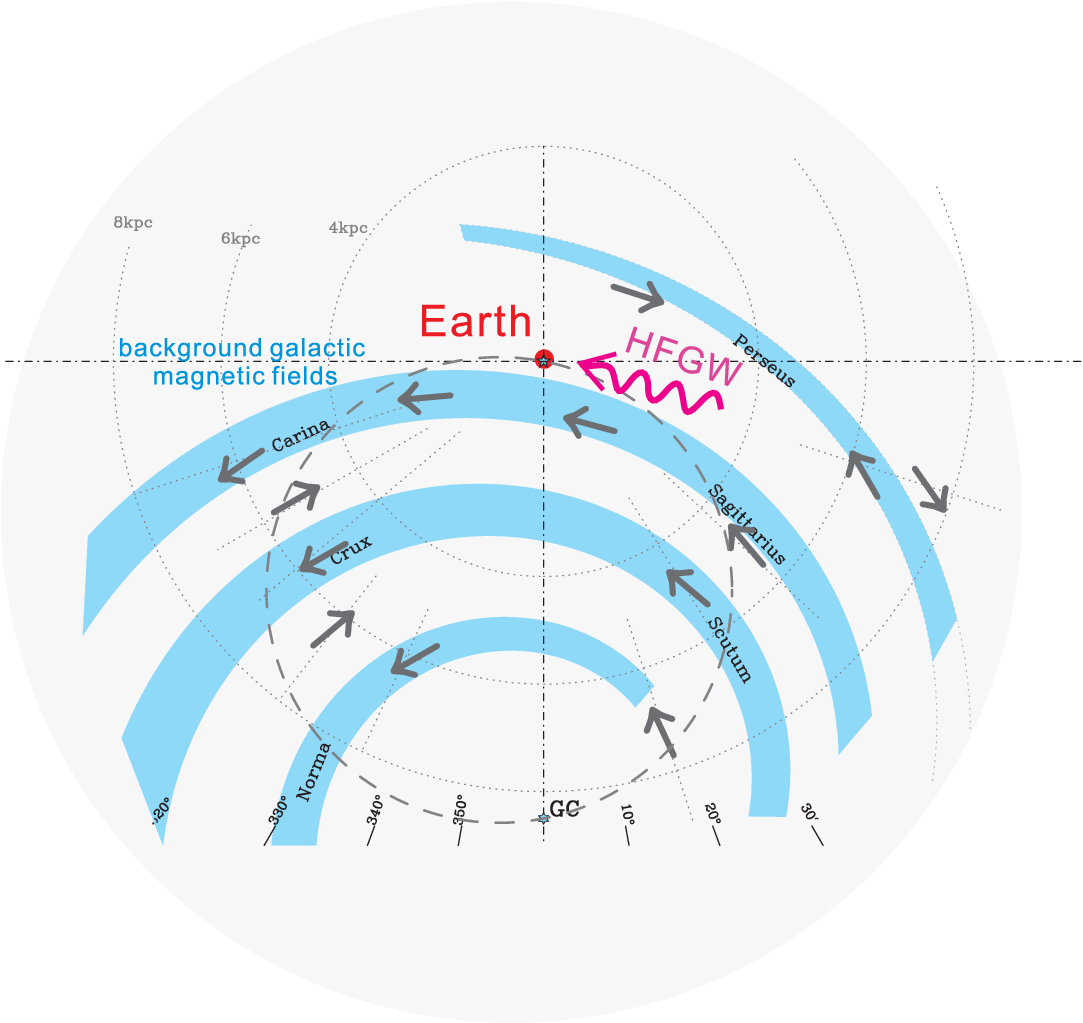}}
	\caption{\footnotesize{   The spatial scale of distribution for the background magnetic fields in the Milky Way reaches up to $\sim$ thousand light-years long\cite{galacticB.RevModPhys.74.775,0004-637X-663-1-258,0004-637X-642-2-868}, and the magnetic fields have a stable strength and direction in the scale. Thus any HFGWs passing through the background magnetic field would generate a significant space accumulation effect in the Earth region. This figure is made based on some materials of the Ref. \cite{0004-637X-642-2-868}.}}
	\label{figure08}
\end{figure}

(4) The EM response to the HFGWs in the transverse background static electric field $\hat E_y^{(0)}$. 
Then  $\hat E_x^{(0)}=\hat E_z^{(0)}=\hat E_x^{(0)}$$=\hat B_y^{(0)}=\hat B_z^{(0)}=0$ in Eqs. (\ref{eq27}) to (\ref{eq31}), i.e., only $\hat E_y^{(0)}$ has non-vanishing value. According to the same way, we find (see Fig. \ref{figure06}):  
\begin{eqnarray}
	\label{eq51}
	n_{z\tiny{\textcircled{1}}}^{(2)}=\dfrac{Re\langle \tilde E^{(1)*}_x \tilde B^{(1)}_y	\rangle}{2\mu_0\hbar\omega_e} = \frac{k_g^2z^2}{8\mu_0c\hbar\omega_e}(A_{\otimes}\hat E_y^{(0)})^2,
\end{eqnarray}
\begin{eqnarray}
	\label{eq52}
	n_{z\tiny{\textcircled{2}}}^{(2)}=\dfrac{Re\langle \tilde E^{(1)*}_y \tilde B^{(1)}_x	\rangle}{2\mu_0\hbar\omega_e} = \frac{k_g^2z^2}{8\mu_0c\hbar\omega_e}[(A_{\oplus}+\frac{\sqrt{2}}{2}A_l)\hat E_y^{(0)}]^2,~~~~~~
\end{eqnarray}
\indent (5) The EM response to the HFGWs in the longitudinal stable EM fields $\hat B_z^{(0)}$ and $\hat E_z^{(0)}$.\\
\indent In fact, whether according to the electrodynamic equations in curved spacetime \cite{nuovo129,prd104008}, or the Feynman perturbation techniques to analyze the conversion of GWs into EM waves (and vice versa) \cite{prd2915}, the GWs (including the HFGWs) in the GR framework (i.e., the GWs having only the $\oplus$-type and the $\otimes$-type polarizations) do not generate any perturbation to the longitudinal static EM fields\cite{nuovo129,prd2915,prd104008}. Unlikely, the HFGWs having the additional polarization states will generate EM perturbations to such EM fields. Thus, their physical behaviors are quite different.\\
\indent Putting   $\hat E_x^{(0)}=\hat E_y^{(0)}$$=\hat B_x^{(0)}=\hat B_y^{(0)}=0$  in Eqs. (\ref{eq27}) to (\ref{eq30}), i.e., only $\hat B_z^{(0)}$ and $\hat E_z^{(0)}$ have non-vanishing values. Then Eqs. (\ref{eq27}) to (\ref{eq30}) are reduced to
\begin{eqnarray}
	\label{eq53}
	&&	\tilde E^{(1)}_x=-\dfrac{i}{2}k_gzh_{x}\hat E_z^{(0)}  -\dfrac{i}{2}k_gczh_{y}\hat B_z^{(0)} \nonumber\\
	&&	=-\dfrac{i}{2}k_gz(A_{x}\hat E_z^{(0)}+cA_y\hat B_z^{(0)})\exp[i(k_gz-\omega_gt)],
\end{eqnarray}
\begin{eqnarray}
	\label{eq54}
	&&	\tilde B^{(1)}_y=-\dfrac{i}{2c}k_gzh_{x}\hat E_z^{(0)}  -\dfrac{i}{2}k_gzh_{y}\hat B_z^{(0)} \nonumber\\
	&&	=-\dfrac{i}{2}k_gz(\frac{1}{c}A_{x}\hat E_z^{(0)}+A_y\hat B_z^{(0)})\exp[i(k_gz-\omega_gt)],
\end{eqnarray}
\begin{eqnarray}
	\label{eq55}
	&&	\tilde E^{(1)}_y=-\dfrac{i}{2}k_gzh_{y}\hat E_z^{(0)} +\dfrac{i}{2}k_gczh_{x}\hat B_z^{(0)} \nonumber\\
	&&	=-\dfrac{i}{2}k_gz(A_{y}\hat E_z^{(0)}-cA_x\hat B_z^{(0)})\exp[i(k_gz-\omega_gt)],
\end{eqnarray}
\begin{eqnarray}
	\label{eq56}
	&&	\tilde B^{(1)}_x=\dfrac{i}{2c}k_gzh_{y}\hat E_z^{(0)}-\dfrac{i}{2}k_gzh_{x}\hat B_z^{(0)} \nonumber\\
	&&	=\dfrac{i}{2}k_gz(\frac{1}{c}A_{y}\hat E_z^{(0)}-A_x\hat B_z^{(0)})\exp[i(k_gz-\omega_gt)],
\end{eqnarray}
From Eqs. (\ref{eq53}) to (\ref{eq56}), the corresponding PPFs can be given by (see Fig. \ref{figure07})
\begin{eqnarray}
	\label{eq57}
	n_{z\tiny{\textcircled{1}}}^{(2)}=\dfrac{Re\langle \tilde E^{(1)*}_x \tilde B^{(1)}_y	\rangle}{2\mu_0\hbar\omega_e} = \frac{k_g^2z^2c}{8\mu_0\hbar\omega_e}(\frac{1}{c}A_{x}\hat E_z^{(0)}+A_{y}\hat B_z^{(0)})^2, ~~~~~
\end{eqnarray}
\begin{eqnarray}
	\label{eq58}
	n_{z\tiny{\textcircled{2}}}^{(2)}=\dfrac{Re\langle \tilde E^{(1)*}_y \tilde B^{(1)}_x	\rangle}{2\mu_0\hbar\omega_e} = \frac{k_g^2z^2c}{8\mu_0\hbar\omega_e}(\frac{1}{c}A_{y}\hat E_z^{(0)}-A_{x}\hat B_z^{(0)})^2,~~~~~
\end{eqnarray} 
\indent It is interesting to note, such PPFs are only produced by the pure additional polarization states (the $x$-type and the $y$-type polarizations, i.e., the vector mode gravitons). In other words, they are independent of the tensor mode gravitons (the $\oplus$-type and the $\otimes$-type polarizations) and the scalar mode gravitons (the $b$-type and the $l$-type polarizations).\\
\indent If we consider only the EM response of the longitudinal stable magnetic field $\hat B_z^{(0)}$, i.e., $\hat E_z^{(0)}=0$ in Eqs. (\ref{eq57}) and (\ref{eq58}), then the equations are reduced to
\begin{eqnarray}
	\label{eq59}
	&&n_{z\tiny{\textcircled{1}}}^{(2)} = \frac{k_g^2z^2c}{8\mu_0\hbar\omega_e}( A_{y}\hat B_z^{(0)})^2,
\end{eqnarray}
\begin{eqnarray}
	\label{eq60}
	&&n_{z\tiny{\textcircled{2}}}^{(2)} = \frac{k_g^2z^2c}{8\mu_0\hbar\omega_e}( A_{x}\hat B_z^{(0)})^2.
\end{eqnarray}
\indent In this case, the $x$-type and the $y$-type polarizations of the HFGWs can be more clearly and directly displayed. In fact, according to contemporary astronomic observation\cite{galacticB.RevModPhys.74.775}, it is certain that there are very widespread background galactic-extragalactic magnetic fields with strengths $\sim10^{-11}T$ to  $10^{-9}T$ within $1Mpc$ in galaxies and galaxy clusters (see Fig. \ref{figure08}). These magnetic fields might provide a large space accumulation effect during the propagating of the HFGWs from possible sources to the Earth. This means that either the EM response to the HFGWs in the transverse background EM fields [see Eqs. (\ref{eq35}), (\ref{eq36}),  (\ref{eq39}), (\ref{eq40}), (\ref{eq45}), (\ref{eq46}), (\ref{eq51}) and (\ref{eq52})] or in the longitudinal background EM fields [see Eqs. (\ref{eq57}) to (\ref{eq60})], it is all possible to detect or observe such effect provided these background EM fields are distributed in very widespread region. Since the wide distribution of the background galactic-extragalactic magnetic fields has been observed by the observational evidence\cite{galacticB.RevModPhys.74.775}, the EM response to the HFGWs in such background magnetic fields would have more realistic significance than that in the background electric fields, and the EM response in the background longitudinal magnetic fields, Eqs. (\ref{eq59}) and (\ref{eq60}), might provide observation evidence produced by the pure-additional polarization states (the $x$- and the $y$-polarizations of the HFGW).\\
\indent It should be pointed out that the coupling between the longitudinal perturbative electric fields $\tilde E_z^{(1)}$, Eq.(\ref{eq31}), and the transverse perturbative magnetic fields, Eqs. (\ref{eq28}) and (\ref{eq30}), do not generate any transverse PPFs, i.e., 
\begin{eqnarray}
	\label{eq61}
	n_{y}^{(2)}=\dfrac{Re\langle \tilde E^{(1)*}_z \tilde B^{(1)}_x	\rangle}{2\mu_0\hbar\omega_e} = 0,\nonumber\\
	n_{x}^{(2)}=\dfrac{Re\langle \tilde E^{(1)*}_z \tilde B^{(1)}_y	\rangle}{2\mu_0\hbar\omega_e} = 0,
\end{eqnarray}
This is because the $\tilde E_z^{(1)}$, Eq. (\ref{eq31}), and $\tilde B_x^{(1)}$, $\tilde B_y^{(1)}$, Eqs. (\ref{eq28}), (\ref{eq30}), have a phase difference of $\pi/2$. Therefore, the average values of the transverse perturbative EM power fluxes, Eq. (\ref{eq61}), are vanishing. In fact, such results ensure the total momentum conservation in the interaction of the HFGWs (gravitons) with the background EM fields. However, the longitudinal perturbative electric field  $\tilde E_z^{(1)}$ will play an important role in the 3DSR system (see below, sect. \ref{3DSR}). We will   show	 that the transverse PPFs generated by the coupling between the  $\tilde E_z^{(1)}$ and the transverse magnetic fields of the Gaussian beam (GB), might display the effect of the pure additional polarization states (the $x$-type, $y$-type, $l$-type and  $b$-type polarizations) of the HFGWs.\\

\section{Electromagnetic counterparts to the HFGWs having the additional polarization states in the 3D-EM synchro-resonance system (3DSR system)}
\label{3DSR}
\begin{figure}[!htbp]
	\centerline{\includegraphics[scale=0.62]{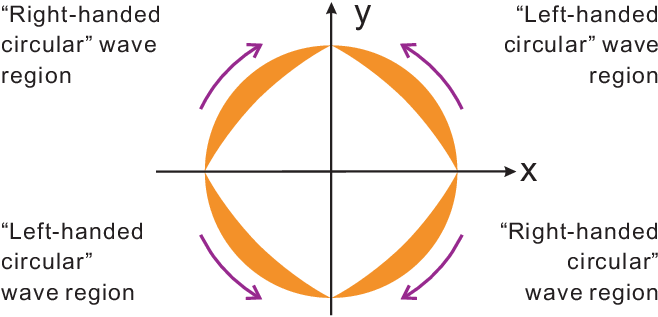}}
	\caption{\footnotesize{The strength distribution of transverse background photon flux $n_{\phi}^{(0)}$, Eq. (\ref{eq64}), in the cylindrical polar coordinates.}}
	\label{figure09}
\end{figure}

\begin{figure*}[t]
	
		\centering
		\subfigure[ ]{
			\begin{minipage}[b]{0.18\linewidth}
				\centering
			\label{fig9a}%
			\includegraphics[width=1.0 in]{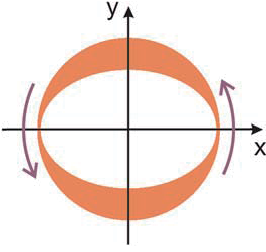}	
		\end{minipage}
    	}
		\subfigure[ ]{
			\begin{minipage}[b]{0.18\linewidth}
				\centering
			\label{fig9b}%
			\includegraphics[width=1.1 in]{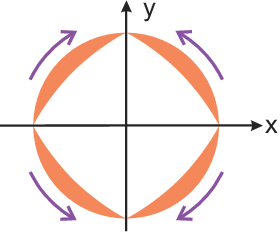}
			\end{minipage}
		}
		\subfigure[ ]{
			\begin{minipage}[b]{0.18\linewidth}
				\centering
			\label{fig9c}%
			\includegraphics[width=1.1 in]{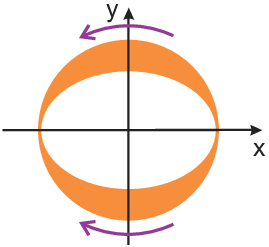}
			\end{minipage}
		}
		\subfigure[ ]{
			\begin{minipage}[b]{0.18\linewidth}
				\centering
			\label{fig9d}%
			\includegraphics[width=1.1 in]{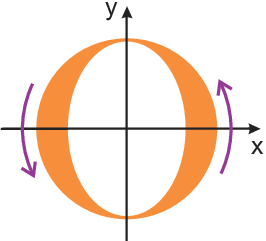}
			\end{minipage}
        }
		\subfigure[ ]{
			\label{fig9e}%
			\includegraphics[width=0.98 in]{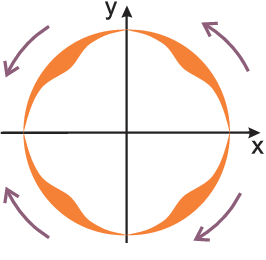}}
		\caption{{The angular distribution (in the cylindrical polar coordinates) of the strengths and ``rotational directions'' of the PPFs, produced by the HFGWs having different polarization states. Here, for subfigures (a) to (e), we have $n^{(1)}_{\phi}\propto\sin^2\phi$,  $n^{(1)}_{\phi}\propto\sin 2\phi$, $n^{(1)}_{\phi}\propto\sin\phi$, $n^{(1)}_{\phi-\oplus,b,l}\propto\cos^2\phi$, $n^{(1)}_{\phi-y}\propto\sin\phi\cos^2\phi$, respectively.}}%
		\label{polarizations}%

\end{figure*}

\indent The 3DSR system consists of the background Gaussian-type photon flux [Gaussian beam (GB)] and the background static EM fields, 
$\hat E^{(0)}$ and  $\hat B^{(0)}$ (see Figs. \ref{figure01} and \ref{figure02}).      The 3DSR was discussed in Ref. \cite{prd104008,FYLi_EPJC_2008}, so we shall not repeat it in detail here.  In this article, the 3DSR system is different to that in previous studies, and we  update it into a new system for probing the HFGWs 
having additional polarizations to display related novel effects, i.e: \\
\indent (1) Unlike previous EM detection schemes, here the 3DSR system contains not only the static magnetic field, but also the static electric field, and their directions can be adjusted. In this special coupling between the static EM fields and the Gaussian-type photon flux, the perturbative EM signals generated by the different polarization states (the tensor, the vector and the scalar mode gravitons in the high-frequency band) can be effectively distinguished and displayed.\\
\indent (2) Since the EM signals generated by the interaction of the HFGWs with the background static EM fields, will have  the same frequencies with the HFGWs. Thus once the GB is adjusted to the resonance frequency band for the HFGWs, then the first-order perturbative EM power fluxes [the perturbative photon fluxes (PPFs), i.e., the signal photon fluxes] also have such frequency band. This means that the 3DSR  can be  a detection system of broad frequency band.\\
\indent In order to make the system having a good sensitivity to distinguish the PPFs generated by the different polarization states of the HFGWs, we select a new group of wave beam solutions for the GB in the framework of the quantum electronic (also see Appendix A):
\begin{eqnarray}
	\label{eq62}
	&&\tilde E_x^{(0)}=\psi_{ex}=\psi=\frac{\psi_0}{\sqrt{1+(z/f)^2}}\cdot\nonumber\\
	&&\exp(\frac{-r^2}{W^2})\exp\{i[(k_e z-\omega _e t)-tan^{-1}\frac{z}{f}+\frac{k_er^2}{2R}+\delta ]\},\nonumber\\
	&&\tilde E_y^{(0)}=\psi_{ey}=0,\nonumber\\
	&&\tilde E_z^{(0)}=\psi_{ez}=2xF_1({\bf{x}},k_e,W)=2rcos\phi F_1({\bf{x}},k_e,W),  ~~~~~~~
\end{eqnarray}
\begin{eqnarray}
	\label{eq63}
	&&\tilde B_x^{(0)}=\psi_{bx}=-\frac{i}{\omega_e}\frac{\partial \psi_{ez}}{\partial y}=\frac{\sin 2\phi}{\omega_e}F_2({\bf{x}},k_e,W),\nonumber\\
	&&\tilde B_y^{(0)}=\psi_{by}=-\frac{i}{\omega_e}(\frac{\partial \psi_{ex}}{\partial z  }-\frac{\partial \psi_{ez}   }{\partial x  })\nonumber\\
	&&=\frac{1}{\omega_e}[F_4({\bf{x}},k_e,W)-i(2F_1({\bf{x}},k_e,W)+\cos^2\phi F_3({\bf{x}},k_e,W))],\nonumber\\
	&&\tilde B_z^{(0)}=\frac{i}{\omega_e}\frac{\partial \psi_{ex}}{\partial y}=-\frac{\sin \phi}{\omega_e}[\frac{k_er}{R}+i\frac{2r}{W_0^2[1+(z/f)^2]}]\psi,
\end{eqnarray}
\indent where $\tilde E_x^{(0)}$, $\tilde E_y^{(0)}$,   $\tilde E_z^{(0)}$, $\tilde B_x^{(0)}$,  $\tilde B_y^{(0)}$ and $\tilde B_z^{(0)}$ are the electric and magnetic components in Cartesian coordinate system for the GB, respectively, and only x-component $\tilde E_x^{(0)}$ of the electric field has a standard form of circular mode of the  fundamental frequency GB\cite{Yariv}. The concrete expressions of functions $F_1$, $F_2$, $F_3$ and $F_4$ can be found in the  Appendix A.\\
\indent In fact, there are different solutions of the wave beam for the Helmholtz equation, and they can be the Gaussian-type wave beams or the quasi-Gaussian-type wave beams. One of reasons of selecting such wave beam solutions, Eqs.(\ref{eq62}) and (\ref{eq63}), is that it will be an optimal coupling between the Gaussian-type photon flux and the background static EM fields, and will make the PPFs (the signal photon fluxes) and the background noise photon fluxes having very different physical behaviors in the special local region. These physical behaviors include the propagating direction, strength distribution, decay rate, wave impedance, etc (see below and Appendix B). Thus, such results will greatly improve the distinguishability between the signal photon fluxes and the background noise photons. Also, they will greatly increase the separability among the tensor mode, the vector mode and the scalar mode gravitons. This is the physical origin of the very low standard quantum limit of the 3DSR system (i.e., the high sensitivity of the 3DSR system, e.g., see Ref \cite{Stephenson}).\\
\indent By using Eqs. (\ref{eq62}) and (\ref{eq63}), the average values of the transverse background photon flux (the Gaussian-type photon flux) with respect to time in cylindrical polar coordinates can be given by
\begin{eqnarray}
	\label{eq64}
	&~&n_{\phi}^{(0)}=-n_x^{(0)}\sin\phi+n_y^{(0)}\cos\phi \nonumber\\
	&&=-\frac{c}{\hbar\omega_e}\langle \mathop{T^{01}}\limits^{(0)} \rangle\sin
	\phi+ \frac{c}{\hbar\omega_e}\langle \mathop{T^{02}}\limits^{(0)} \rangle\cos\phi\nonumber\\
	&=&\frac{-1}{2\mu_0\hbar\omega_e}Re\langle\psi_{ez}^*\psi_{by}\rangle\sin\phi
	+\frac{1}{2\mu_0\hbar\omega_e}Re\langle\psi_{ex}^*\psi_{bz}\rangle\cos\phi\nonumber\\
	&+&\frac{1}{2\mu_0\hbar\omega_e}Re\langle\psi_{ez}^*\psi_{bx}\rangle\cos\phi
	=f_{\phi}^{(0)}\exp(-\frac{2r^2}{W^2})\sin2\phi, ~~~~~~~~
\end{eqnarray} 
where  $\mathop{T^{01}}\limits^{(0)}$ and $\mathop{T^{02}}\limits^{(0)}$ are 01- and 02-components of the energy-momentum
tensor for the background EM wave (the GB), and\cite{Yariv}
\begin{eqnarray}
	\label{eq65}
	n_{\phi}^{(0)}|_{x=0}=n_{\phi}^{(0)}|_{y=0}=0.
\end{eqnarray}
i.e., the transverse background photon flux at the longitudinal symmetry surface (the yz-plane and the xz-plane) of the GB is equal to zero  (see Fig. \ref{figure09}). In fact, this is the necessary condition for the stability of  GB.\\
\indent From Eqs. (\ref{eq62}), (\ref{eq63}) and (\ref{eq27}) to (\ref{eq31}), and under the resonance condition ($\omega_e=\omega_g$), the transverse perturbative photon fluxes (PPFs) in cylindrical polar coordinates can be given by
\begin{eqnarray}
	\label{eq66}
	&~&n_{\phi}^{(1)}=-n_{x}^{(1)}\sin\phi+n_{y}^{(1)}\cos\phi
	=-\frac{c}{2\hbar\omega_e}\nonumber\\
	&&\cdot\langle \mathop{T^{01}}\limits^{(1)} \rangle_{\omega_e=\omega_g}\sin
	\phi+ \frac{c}{2\hbar\omega_e}\langle \mathop{T^{02}}\limits^{(1)} \rangle_{\omega_e=\omega_g}\cos\phi\nonumber\\
	&=&-\frac{1}{2\mu_0\hbar\omega_e}Re\langle{\tilde{E}}_y^{(1)*}{\tilde{B}}_{z}^{(0)}\rangle_{\omega_e=\omega_g}\sin\phi\nonumber\\
	&&	+\frac{1}{2\mu_0\hbar\omega_e}Re\langle{\tilde{E}}_z^{(1)*}{\tilde{B}}_{x}^{(0)}\rangle_{\omega_e=\omega_g}\cos\phi\nonumber\\
	&-&\frac{1}{2\mu_0\hbar\omega_e}Re\langle{\tilde{E}}_x^{(1)*}{\tilde{B}}_{z}^{(0)}\rangle_{\omega_e=\omega_g}\cos\phi\nonumber\\
	&&+\frac{1}{2\mu_0\hbar\omega_e}Re\langle{\tilde{E}}_z^{(1)*}{\tilde{B}}_{x}^{(0)}\rangle_{\omega_e=\omega_g}\cos\phi,
\end{eqnarray}
where $\langle \mathop{T^{01}}\limits^{(1)} \rangle_{\omega_e=\omega_g}$ and
$\langle \mathop{T^{02}}\limits^{(1)} \rangle_{\omega_e=\omega_g}$ are  average values with respect to time of  01- and 02-components of energy-momentum tensor for the first-order perturbative EM fields.\\
\indent In the following, we shall study the EM response to the HFGWs with the additional polarization states in some of typical cases.\\
\indent 1. EM response to the HFGWs in the coupling system between the transverse background static magnetic field $\hat B_y^{(0)}$ and the GB.\\
\indent Then $\hat E_x^{(0)}=\hat E_y^{(0)}=\hat E_z^{(0)} =\hat B_x^{(0)}=\hat B_z^{(0)}=0$  and only $\hat B_y^{(0)}\neq 0$. In fact, this is a coupling system between the transverse static magnetic fields $\tilde{B}_y^{(0)}$  and the Gaussian-type photon flux. In this case, from Eqs. (\ref{eq33}) to (\ref{eq34}) and  (\ref{eq62}), (\ref{eq63}) and (\ref{eq66}), the concrete forms of the transverse PPFs can be obtained:
\begin{spacing}{1.0}
	\begin{widetext}
		\begin{eqnarray}
			\label{eq67}
			n_{\phi-\otimes}^{(1)}&=&\frac{-1}{2\mu_0\hbar\omega_e}Re\langle{\tilde{E}}_y^{(1)*}{\tilde{B}}_{z}^{(0)}\rangle_{\omega_e=\omega_g}\sin\phi 
			=\frac{1}{\mu_0\hbar\omega_e}\{
			\frac{A_{\otimes}\hat{B}_y^{(0)}\psi_0k_g\Delta \textit{z}~r}{2[1+(z/f)^2]^{\frac{1}{2}}(z+f^2/z)}
			\sin[\frac{k_e r^2}{2R} -\tan ^{-1}(\frac{z}{f})+\delta]\nonumber\\
			&+&\frac{A_{\otimes}\hat{B}_y^{(0)}
				\psi_0\Delta\textit{z}~r}{W_0^2[1+(z/f)^2]^{\frac{3}{2}}}
			\cos[\frac{k_e r^2}{2R} -\tan ^{-1}(\frac{z}{f})+\delta]\}exp(-\frac{r^2}{W^2})\sin^2\phi,
		\end{eqnarray}
		\begin{eqnarray}
			\label{eq68}
			n_{\phi-\oplus,l}^{(1)}&=&\frac{1}{2\mu_0\hbar\omega_e}Re\langle{\tilde{E}}_x^{(1)*}{\tilde{B}}_{z}^{(0)}\rangle_{\omega_e=\omega_g}\cos\phi=\frac{1}{\mu_0\hbar\omega_e}\{
			\frac{(A_{\oplus}+\frac{\sqrt{2}}{2}A_l)\hat{B}_y^{(0)}\psi_0k_g\Delta \textit{z}~r}{4[1+(z/f)^2]^{\frac{1}{2}}(z+f^2/z)}
			\sin[\frac{k_e r^2}{2R} -\tan ^{-1}(\frac{z}{f})+\delta]\nonumber\\
			&+&\frac{(A_{\oplus}+\frac{\sqrt{2}}{2}A_l)\hat{B}_y^{(0)}
				\psi_0\Delta\textit{z}~r}{2W_0^2[1+(z/f)^2]^{\frac{3}{2}}}
			\cos[\frac{k_e r^2}{2R} -\tan ^{-1}(\frac{z}{f})+\delta]\}exp(-\frac{r^2}{W^2})\sin2\phi,
		\end{eqnarray}
	\end{widetext}
\end{spacing}
where $\Delta z$ is the spatial scale of the transverse static magnetic field $\hat B_y^{(0)}$ in the 3DSR system.
Eq. (\ref{eq67}) shows that the transverse PPF $n_{\phi-\otimes}^{(1)}$ is produced by the pure $\otimes$-type polarization state of the HFGWs, while Eq. (\ref{eq68}) represents that the transverse PPF $n_{\phi-\oplus,l}^{(1)}$ is generated by the combination state of the $\oplus$-type and the $l$-type polarizations of the HFGWs.\\
\indent 2. EM response to the HFGWs in the coupling system between the transverse background static magnetic field $\hat B_x^{(0)}$ and the GB.\\
\indent Then $\hat E_x^{(0)}=\hat E_y^{(0)}=\hat E_z^{(0)} =\hat B_y^{(0)}=\hat B_z^{(0)}=0$  and only $\hat B_x^{(0)}\neq 0$. In this case, from Eqs.  (\ref{eq63}) and (\ref{eq66}), we have,
\begin{widetext}
\begin{eqnarray}
	\label{eq69}
	n_{\phi-\oplus,b,l}^{(1)}&=&\frac{-1}{\mu_0\hbar\omega_e}Re\langle{\tilde{E}}_y^{(1)*}{\tilde{B}}_{z}^{(0)}\rangle_{\omega_e=\omega_g}\sin\phi=\frac{1}{\mu_0\hbar\omega_e}
	\cdot\{
	\frac{(A_{b}-A_{\oplus}+\sqrt{2}A_l)\hat{B}_x^{(0)}\psi_0k_g\Delta \textit{z}~r}{2[1+(z/f)^2]^{\frac{1}{2}}(z+f^2/z)}
	\sin[\frac{k_e r^2}{2R} -\tan ^{-1}(\frac{z}{f})+\delta]\nonumber\\
	&+&\frac{(A_{b}-A_{\oplus}+\sqrt{2}A_l)\hat{B}_x^{(0)}
		\psi_0\Delta\textit{z}~r}{2W_0^2[1+(z/f)^2]^{\frac{3}{2}}}
	\cos[\frac{k_e r^2}{2R} -\tan ^{-1}(\frac{z}{f})+\delta]\}
	\cdot \exp(-\frac{r^2}{W^2})\sin^2\phi, 
\end{eqnarray}
\begin{eqnarray}
	\label{eq70}
	n_{\phi-\otimes}^{(1)}&=&\frac{1}{\mu_0\hbar\omega_e}Re\langle{\tilde{E}}_x^{(1)*}{\tilde{B}}_{z}^{(0)}\rangle_{\omega_e=\omega_g}\cos\phi=\frac{1}{\mu_0\hbar\omega_e}
	\cdot\{
	\frac{A_{\otimes}\hat{B}_x^{(0)}\psi_0k_g\Delta \textit{z}~r}{4[1+(z/f)^2]^{\frac{1}{2}}(z+f^2/z)}
	\sin[\frac{k_e r^2}{2R} -\tan ^{-1}(\frac{z}{f})+\delta]\nonumber\\
	&+&\frac{A_{\otimes}\hat{B}_x^{(0)}
		\psi_0\Delta\textit{z}~r}{2W_0^2[1+(z/f)^2]^{\frac{3}{2}}}
	\cos[\frac{k_e r^2}{2R} -\tan ^{-1}(\frac{z}{f})+\delta]\}\exp(-\frac{r^2}{W^2})\sin 2\phi, 
\end{eqnarray}
\end{widetext}
Eq. (\ref{eq69}) shows that the  PPF $n_{\phi-\oplus,b,l}^{(1)}$ is produced by the combination state of the $\oplus$-type, $b$-type and $l$-type polarizations of the HFGWs,   while Eq. (\ref{eq70}) represents that the PPF $n_{\phi-\otimes}^{(1)}$ is generated by the pure   $\otimes$-type   polarization of the HFGWs.\\
\indent 3. The EM response to the HFGWs in the coupling system between the longitudinal background static magnetic field $\hat B_z^{(0)}$ and the GB.\\
\indent Then $\hat E_x^{(0)}=\hat E_y^{(0)}=\hat E_z^{(0)} =\hat B_x^{(0)}=\hat B_y^{(0)}=0$  and only $\hat B_z^{(0)}\neq 0$.  In this case, under the resonance condition ($\omega_e=\omega_g$), from Eqs. (\ref{eq62}), (\ref{eq63}) and (\ref{eq66}), the concrete forms of the transverse PPFs can be given by:\\
\begin{widetext}
	\begin{eqnarray}
		\label{eq75}
		n_{\phi-x}^{(1)}&=&\frac{-1}{\mu_0\hbar\omega_e}Re\langle{\tilde{E}}_y^{(1)*}{\tilde{B}}_{z}^{(0)}\rangle_{\omega_e=\omega_g}\sin\phi=\frac{1}{\mu_0\hbar\omega_e}\{
		\frac{A_{x}\hat{B}_z^{(0)}\psi_0k_g\Delta \textit{z}~r}{2[1+(z/f)^2]^{\frac{1}{2}}(z+f^2/z)}
		\sin[\frac{k_e r^2}{2R} -\tan ^{-1}(\frac{z}{f})+\delta]\nonumber\\
		&+&\frac{A_{x}\hat{B}_z^{(0)}
			\psi_0\Delta\textit{z}~r}{W_0^2[1+(z/f)^2]^{\frac{3}{2}}}
		\cos[\frac{k_e r^2}{2R} -\tan ^{-1}(\frac{z}{f})+\delta]\}\exp(-\frac{r^2}{W^2})\sin^2\phi,
	\end{eqnarray}
	\begin{eqnarray}
		\label{eq76}
		n_{\phi-y}^{(1)}&=&\frac{1}{\mu_0\hbar\omega_e}Re\langle{\tilde{E}}_x^{(1)*}{\tilde{B}}_{z}^{(0)}\rangle_{\omega_e=\omega_g}\cos\phi=\frac{1}{\mu_0\hbar\omega_e}\{
		\frac{A_{y}\hat{B}_z^{(0)}\psi_0k_g\Delta \textit{z}~r}{4[1+(z/f)^2]^{\frac{1}{2}}(z+f^2/z)}
		\sin[\frac{k_e r^2}{2R} -\tan ^{-1}(\frac{z}{f})+\delta]\nonumber\\
		&+&\frac{A_{y}\hat{B}_z^{(0)}
			\psi_0\Delta\textit{z}~r}{2W_0^2[1+(z/f)^2]^{\frac{3}{2}}}
		\cos[\frac{k_e r^2}{2R} -\tan ^{-1}(\frac{z}{f})+\delta]\}\exp(-\frac{r^2}{W^2})\sin 2\phi,
	\end{eqnarray}
\end{widetext}
Eqs. (\ref{eq75}) and (\ref{eq76}) show that the transverse  PPFs $n_{\phi-x}^{(1)}$ and $n_{\phi-y}^{(1)}$ are generated by the pure  $x$-type and the pure  $y$-type   polarizations of the HFGWs, respectively. \\
\indent 4. The EM response to the HFGWs in the coupling system between the transverse background static electric field $\hat E_x^{(0)}$ and the GB.\\
\indent Then $ \hat E_y^{(0)}=\hat E_z^{(0)} =\hat B_x^{(0)}=\hat B_y^{(0)}=\hat B_z^{(0)}=0$  and only $\hat E_x^{(0)}\neq 0$. In this case  from Eqs. (\ref{eq31}),   (\ref{eq63}) and (\ref{eq66}), in the same way, under the resonance condition ($\omega_e=\omega_g$), the transverse PPFs can be given by
\begin{spacing}{1.0}
	\begin{widetext}
		\begin{eqnarray}
			n_{\phi-\oplus,l}^{(1)}&=&\frac{-1}{2\mu_0\hbar\omega_e}Re\langle{\tilde{E}}_x^{(1)*}{\tilde{B}}_{z}^{(0)}\rangle_{\omega_e=\omega_g}\cos\phi=\frac{1}{\mu_0\hbar\omega_e}\{
			\frac{( A_{\oplus}-\frac{\sqrt{2}}{2}A_l)\hat{E}_x^{(0)}\psi_0k_g\Delta \textit{z}~r}{4[1+(z/f)^2]^{\frac{1}{2}}(z+f^2/z)}
			\sin[\frac{k_e r^2}{2R} -\tan ^{-1}(\frac{z}{f})+\delta]\nonumber
		\end{eqnarray}
		\begin{eqnarray}
			\label{eq77}
			&+&\frac{(A_{\oplus}-\frac{\sqrt{2}}{2}A_l)\hat{E}_x^{(0)}
				\psi_0\Delta\textit{z}~r}{2W_0^2[1+(z/f)^2]^{\frac{3}{2}}}
			\cos[\frac{k_e r^2}{2R} -\tan ^{-1}(\frac{z}{f})+\delta]\}\exp(-\frac{r^2}{W^2})\sin 2\phi,
		\end{eqnarray}
		\begin{subequations}\label{eq78}
			\begin{eqnarray}
				\label{eq78a}
				n_{\phi-x}^{(1)}&=&\frac{-1}{2\mu_0\hbar\omega_e}Re\langle{\tilde{E}}_y^{(1)*}{\tilde{B}}_{z}^{(0)}\rangle_{\omega_e=\omega_g}\sin\phi
				+\frac{1}{2\mu_0\hbar\omega_e}Re\langle{\tilde{E}}_z^{(1)*}{\tilde{B}}_{x}^{(0)}\rangle_{\omega_e=\omega_g}\cos\phi\nonumber\\
				&+&\frac{1}{2\mu_0\hbar\omega_e}Re\langle{\tilde{E}}_z^{(1)*}{\tilde{B}}_{y}^{(0)}\rangle_{\omega_e=\omega_g}\sin\phi=\frac{1}{\mu_0\hbar\omega_e}\{
				\frac{A_{x}\hat{E}_x^{(0)}\psi_0k_g\Delta \textit{z}~r}{2[1+(z/f)^2]^{\frac{1}{2}}(z+f^2/z)}
				\sin[\frac{k_e r^2}{2R} -\tan ^{-1}(\frac{z}{f})+\delta]\nonumber\\
				&+&\frac{A_{x}\hat{E}_x^{(0)}
					\psi_0\Delta\textit{z}~r}{W_0^2[1+(z/f)^2]^{\frac{3}{2}}}
				\cos[\frac{k_e r^2}{2R} -\tan ^{-1}(\frac{z}{f})+\delta]\}\exp(-\frac{r^2}{W^2})\sin^2\phi,
			\end{eqnarray}
			\begin{eqnarray}
				\label{eq78b}
				+\frac{1}{\mu_0\hbar\omega_e}Re\{A_x\hat{E}_x^{(0)}Re\langle\frac{1}{\omega_e}\exp[i(k_gz-\omega_gt)]^*\cdot F_2 \rangle_{\omega_e=\omega_g}\}\sin\phi\cos^2\phi
			\end{eqnarray}
			\begin{eqnarray}
				\label{eq78c}
				+\frac{1}{\mu_0\hbar\omega_e}Re\{\frac{A_x\hat{E}_x^{(0)}}{2}Re\langle\frac{1}{\omega_e}\exp[i(k_gz-\omega_gt)]^*\cdot (F_4-2iF_1) \rangle_{\omega_e=\omega_g}\}\sin\phi
			\end{eqnarray}
			\begin{eqnarray}
				\label{eq78d}
				-\frac{i}{\mu_0\hbar\omega_e}Re\{\frac{A_x\hat{E}_x^{(0)}}{2}Re\langle\frac{1}{\omega_e}\exp[i(k_gz-\omega_gt)]^*\cdot F_3 \rangle_{\omega_e=\omega_g}\}\sin\phi\cos^2\phi
			\end{eqnarray}
		\end{subequations}
		Eqs. (\ref{eq77}) and (\ref{eq78}) show that the transverse  PPFs $n_{\phi-\oplus,l}^{(1)}$ is generated by the combination state of the $\oplus$-type and the $l$-type polarizations, and $n_{\phi-x}^{(1)}$ is produced by the pure $x$-type polarization state.\\
		\indent 5. The EM response to the HFGWs in the coupling system between the transverse background static electric field $\hat E_y^{(0)}$ and the GB.\\
		\indent Then $ \hat E_x^{(0)}=\hat E_z^{(0)} =\hat B_x^{(0)}=\hat B_y^{(0)}=\hat B_z^{(0)}=0$  and only $\hat E_y^{(0)}\neq 0$. In this case  from Eqs. (\ref{eq31}),   (\ref{eq63}) and (\ref{eq66}), in the same way, under the resonance condition ($\omega_e=\omega_g$), the transverse PPFs can be given by
		\begin{eqnarray}
			\label{eq79}
			n_{\phi-\otimes}^{(1)}&=&\frac{-1}{2\mu_0\hbar\omega_e}Re\langle{\tilde{E}}_x^{(1)*}{\tilde{B}}_{z}^{(0)}\rangle_{\omega_e=\omega_g}\cos\phi
			=\frac{1}{\mu_0\hbar\omega_e}\{
			\frac{A_{\otimes}\hat{E}_y^{(0)}\psi_0k_g\Delta \textit{z}~r}{4[1+(z/f)^2]^{\frac{1}{2}}(z+f^2/z)}
			\sin[\frac{k_e r^2}{2R} -\tan ^{-1}(\frac{z}{f})+\delta]\nonumber\\
			&+&\frac{A_{\otimes}\hat{E}_y^{(0)}
				\psi_0\Delta\textit{z}~r}{2W_0^2[1+(z/f)^2]^{\frac{3}{2}}}
			\cos[\frac{k_e r^2}{2R} -\tan ^{-1}(\frac{z}{f})+\delta]\}\exp(-\frac{r^2}{W^2})\sin 2\phi,
		\end{eqnarray}
		\begin{eqnarray}
			\label{eq80}
			n_{\phi-\oplus,l}^{(1)}&=&\frac{-1}{2\mu_0\hbar\omega_e}Re\langle{\tilde{E}}_y^{(1)*}{\tilde{B}}_{z}^{(0)}\rangle_{\omega_e=\omega_g}\sin\phi
			=\frac{1}{\mu_0\hbar\omega_e}\{
			\frac{( A_{\oplus}+\frac{\sqrt{2}}{2}A_l)\hat{E}_y^{(0)}\psi_0k_g\Delta \textit{z}~r}{2[1+(z/f)^2]^{\frac{1}{2}}(z+f^2/z)}
			\sin[\frac{k_e r^2}{2R} -\tan ^{-1}(\frac{z}{f})+\delta]\nonumber\\
			&+&\frac{(A_{\oplus}+\frac{\sqrt{2}}{2}A_l)\hat{E}_y^{(0)}
				\psi_0\Delta\textit{z}~r}{W_0^2[1+(z/f)^2]^{\frac{3}{2}}}
			\cos[\frac{k_e r^2}{2R} -\tan ^{-1}(\frac{z}{f})+\delta]\}\exp(-\frac{r^2}{W^2})\sin^2\phi,
		\end{eqnarray}
		\begin{subequations}\label{eq81}
			\begin{eqnarray}
				\label{eq81a}
				n_{\phi-y}^{(1)}&=&\frac{1}{2\mu_0\hbar\omega_e}Re\langle{\tilde{E}}_z^{(1)*}{\tilde{B}}_{x}^{(0)}\rangle_{\omega_e=\omega_g}\cos\phi
				+\frac{1}{2\mu_0\hbar\omega_e}Re\langle{\tilde{E}}_z^{(1)*}{\tilde{B}}_{y}^{(0)}\rangle_{\omega_e=\omega_g}\sin\phi\nonumber\\
				&=&\frac{1}{\mu_0\hbar\omega_e}Re\{A_y\hat{E}_y^{(0)}Re\langle\frac{1}{\omega_e}\exp[i(k_gz-\omega_gt)]^*\cdot F_2 \rangle_{\omega_e=\omega_g}\}\sin\phi\cos^2\phi
			\end{eqnarray}
			\begin{eqnarray}
				\label{eq81b}
				+\frac{1}{\mu_0\hbar\omega_e}Re\{\frac{A_y\hat{E}_y^{(0)}}{2}Re\langle\frac{1}{\omega_e}\exp[i(k_gz-\omega_gt)]^*\cdot (F_4-2iF_1) \rangle_{\omega_e=\omega_g}\}\sin\phi
			\end{eqnarray}
			\begin{eqnarray}
				\label{eq81c}
				-\frac{i}{\mu_0\hbar\omega_e}Re\{\frac{A_y\hat{E}_y^{(0)}}{2}Re\langle\frac{1}{\omega_e}\exp[i(k_gz-\omega_gt)]^*\cdot F_3 \rangle_{\omega_e=\omega_g}\}\sin\phi\cos^2\phi
			\end{eqnarray}
		\end{subequations}	
	\end{widetext}
\end{spacing}
Eq. (\ref{eq79}) shows that the transverse PPF $n_{\phi-\otimes}^{(1)}$ is generated by the pure $\otimes$-type polarization state; Eq. (\ref{eq80}) shows that the transverse PPF $n_{\phi-\oplus,l}^{(1)}$ is generated by the combination state of the $\oplus$-type and the $l$-type polarizations, and Eq. (\ref{eq81}) shows that $n_{\phi-y}^{(1)}$ is produced by the pure $y$-type polarization state.\\
\indent 6. The EM response to the HFGWs in the coupling system between the longitudinal background static electric field $\hat E_z^{(0)}$ and the GB.\\
\indent Then $ \hat E_x^{(0)}=\hat E_y^{(0)} =\hat B_x^{(0)}=\hat B_y^{(0)}=\hat B_z^{(0)}=0$  and only $\hat E_z^{(0)}\neq 0$.
In the same way, under the resonance condition ($\omega_e=\omega_g$), the transverse PPFs, can be given by
\begin{spacing}{1.0}
	\begin{widetext}
		\begin{eqnarray}
			\label{eq87}
			n_{\phi-x}^{(1)}&=&\frac{-1}{2\mu_0\hbar\omega_e}Re\langle{\tilde{E}}_x^{(1)*}{\tilde{B}}_{z}^{(0)}\rangle_{\omega_e=\omega_g}\cos\phi
			=\frac{1}{\mu_0\hbar\omega_e}\{
			\frac{A_{x}\hat{E}_z^{(0)}\psi_0k_g\Delta \textit{z}~r}{4[1+(z/f)^2]^{\frac{1}{2}}(z+f^2/z)}
			\sin[\frac{k_e r^2}{2R} -\tan ^{-1}(\frac{z}{f})+\delta]\nonumber\\
			&+&\frac{A_{x}\hat{E}_z^{(0)}
				\psi_0\Delta\textit{z}~r}{W_0^2[1+(z/f)^2]^{\frac{3}{2}}}
			\cos[\frac{k_e r^2}{2R} -\tan ^{-1}(\frac{z}{f})+\delta]\}\exp(-\frac{r^2}{W^2})\sin 2\phi,
		\end{eqnarray}
		\begin{eqnarray}
			\label{eq88}
			n_{\phi-y}^{(1)}&=&\frac{-1}{2\mu_0\hbar\omega_e}Re\langle{\tilde{E}}_y^{(1)*}{\tilde{B}}_{z}^{(0)}\rangle_{\omega_e=\omega_g}\sin\phi
			=\frac{1}{\mu_0\hbar\omega_e}\{
			\frac{  A_{y}\hat{E}_z^{(0)}\psi_0k_g\Delta \textit{z}~r}{2[1+(z/f)^2]^{\frac{1}{2}}(z+f^2/z)}
			\sin[\frac{k_e r^2}{2R} -\tan ^{-1}(\frac{z}{f})+\delta]\nonumber\\
			&+&\frac{A_{y}\hat{E}_z^{(0)}
				\psi_0\Delta\textit{z}~r}{W_0^2[1+(z/f)^2]^{\frac{3}{2}}}
			\cos[\frac{k_e r^2}{2R} -\tan ^{-1}(\frac{z}{f})+\delta]\}\exp(-\frac{r^2}{W^2})\sin^2\phi,
		\end{eqnarray}
		\begin{subequations}\label{eq89}
			\begin{eqnarray}
				\label{eq89a}
				n_{\phi-b,l}^{(1)}&=&\frac{1}{2\mu_0\hbar\omega_e}Re\langle{\tilde{E}}_z^{(1)*}{\tilde{B}}_{x}^{(0)}\rangle_{\omega_e=\omega_g}\cos\phi
				+\frac{1}{2\mu_0\hbar\omega_e}Re\langle{\tilde{E}}_z^{(1)*}{\tilde{B}}_{y}^{(0)}\rangle_{\omega_e=\omega_g}\sin\phi\nonumber\\
				&=&\frac{1}{\mu_0\hbar\omega_e}Re\{(\frac{\sqrt{2}}{2} A_l-A_b)\hat{E}_z^{(0)}Re\langle\frac{1}{\omega_e}\exp[i(k_gz-\omega_gt)]^*\cdot F_2 \rangle_{\omega_e=\omega_g}\}\sin\phi\cos^2\phi  
			\end{eqnarray}
			\begin{eqnarray}
				\label{eq89b}
				+\frac{1}{\mu_0\hbar\omega_e}Re\{(\frac{\sqrt{2}}{2} A_l-A_b)\hat{E}_z^{(0)}Re\langle\frac{1}{\omega_e}\exp[i(k_gz-\omega_gt)]^*\cdot (F_4-2iF_1) \rangle_{\omega_e=\omega_g}\}\sin\phi
			\end{eqnarray}
			\begin{eqnarray}
				\label{eq89c}
				-\frac{i}{\mu_0\hbar\omega_e}Re\{(\frac{\sqrt{2}}{2} A_l-A_b)\hat{E}_z^{(0)}Re\langle\frac{1}{\omega_e}\exp[i(k_gz-\omega_gt)]^*\cdot F_3 \rangle_{\omega_e=\omega_g}\}\sin\phi\cos^2\phi
			\end{eqnarray}
		\end{subequations}	
	\end{widetext}
\end{spacing}
Eqs. (\ref{eq87}) and (\ref{eq88})  show that the transverse PPFs, $n_{\phi-x}^{(1)}$ and $n_{\phi-y}^{(1)}$ are generated by the pure $x$-type polarization and the pure $y$-type polarization of the HFGWs, respectively.  The Eq. (\ref{eq89}) shows that the PPF $n_{\phi-b,l}^{(1)}$ is produced by the combination state of the $b$-type and the $l$-type polarizations.\\ 
\indent In all of the above discussions, the ratio [of the electric component ($\tilde E_x^{(1)}$, $\tilde E_y^{(1)}$, $\tilde E_z^{(1)}$) to related magnetic components ($\tilde B_x^{(0)}$, $\tilde B_y^{(0)}$, $\tilde B_z^{(0)}$)] of the PPFs is much less than the ratio of the background noise photon flux. This means that the PPFs expressed by the Eqs. (\ref{eq67}) to (\ref{eq70}), (\ref{eq75}) to (\ref{eq81}), (\ref{eq87}) to (\ref{eq89}) have very low wave impedance\cite{LiNPB2016,Haslett}, which is much less than the wave impedance to the BPFs (see below). Then the PPFs (i.e., the signal photon fluxes) would be easier to pass through the transmission way of the 3DSR system than the BPFs due to very small Ohm losses of the PPFs.\\
\indent According to the same way, and from Eqs. (\ref{eq30}) and (\ref{eq62}), the PPF in the EM response to the HFGWs in the coupling system between the transverse background static magnetic field  $\hat B_x^{(0)}$ and the GB can be given by
\begin{eqnarray}
	\label{eq90}
	&&	n_{\phi-\oplus,b,l}^{(1)}=\frac{1}{2\mu_0\hbar\omega_e}Re\langle{\tilde{E}}_z^{(0)*}{\tilde{B}}_{x}^{(1)}\rangle_{\omega_e=\omega_g}\cos\phi
	\nonumber\\
	&&=\frac{(A_b-A_{\oplus}+\sqrt{2}A_l)\hat{B}_{x}^{(0)}k_g\Delta \textit{z}~r}{2\mu_0\hbar\omega_e}\cdot  \nonumber\\
	&& Re\langle F_1^*\cdot \exp[i(k_gz-\omega_gt)]  \rangle_{\omega_e=\omega_g}\cos^2\phi.
\end{eqnarray}
Eq. (\ref{eq90}) shows that the PPF $n_{\phi-\oplus,b,l}^{(1)}$, is generated by the combination state among the $\oplus$-type, the $b$-type and the $l$-type polarizations of the HFGWs. Obviously the ratio of the electric component to the magnetic component of the PPF is larger than that of the PPF $n_{\phi-\oplus,b,l}^{(1)}$ [Eq. (\ref{eq69})], i.e., the PPF expressed by Eq. (\ref{eq90}) has larger wave impedance. However, its angular distribution factor $\cos^2\phi$ (see Fig. \ref{fig9d}) is quite different to the $\sin 2\phi$ of the BPF, Eq. (\ref{eq64}), Fig. \ref{figure09}. Thus, it is always possible to distinguish them.\\
\indent 7. Angular distributions of the strengths and the ``rotation directions'' of typical PPFs in the cylindrical polar coordinates.\\
\indent Based on the above discussion, the angular distributions of the strengths and the ``rotation directions'' for some of typical transverse PPFs are listed in the following figures [Figs. (\ref{fig9a}) to (\ref{fig9e})].\\

\indent In Fig.  \ref{fig9a}, the $n^{(1)}_{\phi}\propto\sin^2\phi$, and it includes following five cases: \\
(i) $n^{(1)}_{\phi-\otimes}$, Eq. (\ref{eq67}). This is the transverse PPF displaying the pure $\otimes$-type polarization state (the tensor mode gravitons) of the HFGWs. The PPF is from the EM response to the HFGWs in the coupling between the transverse static magnetic field $\hat B_y^{(0)}$ and the GB in the 3DSR.\\
(ii) $n^{(1)}_{\phi-\oplus,b,l}$, Eq. (\ref{eq69}). This is the transverse PPF displaying the combination state of the  $\oplus$-type, the $b$-type and the $l$-type polarizations (the tensor-mode and the scalar-mode gravitons) of the HFGWs. The PPF is from the EM response to the HFGWs in the coupling between the transverse static magnetic field  $\hat B_x^{(0)}$   and the GB in the 3DSR.\\
(iii) $n^{(1)}_{\phi-x}$, Eqs. (\ref{eq75}) and (\ref{eq78a}). This is the transverse PPF displaying the  pure $x$-type polarization state (the vector mode gravitons) of the HFGWs. The PPF is from the EM response to the HFGWs in the coupling between the longitudinal static magnetic field $\hat B_z^{(0)}$ (or the transverse static electric field $\hat E_x^{(0)}$) and the GB in the 3DSR.\\
(iv) $n^{(1)}_{\phi-\oplus,l}$, Eq. (\ref{eq80}). This is the transverse PPF displaying the combination state of the  $\oplus$-type and the $l$-type polarizations (the tensor-mode and the scalar-mode gravitons) of the HFGWs. The PPF is from the EM response to the HFGWs in the coupling between the transverse static electric field  $\hat E_y^{(0)}$   and the GB in the 3DSR.\\
(v) $n^{(1)}_{\phi-y}$, Eq. (\ref{eq88}). This is the transverse PPF displaying the pure $y$-type polarization state (the vector mode gravitons) of the HFGWs. The PPF is from the EM response to the HFGWs in the coupling between the longitudinal static electric field $\hat E_z^{(0)}$ and the GB in the 3DSR.\\
\indent Best detection position of all of such PPFs should be the receiving surfaces at $\phi=\pi/2$ and $\phi=3\pi/2$ (see Fig. \ref{fig9a}), where the PPFs have their maximum, while the BPF (the background noise photon flux) vanishes at the surface (see Fig.  \ref{figure09}). Because the BPF from the GB will be the dominant source of the noise photon fluxes, i.e.,  other noise photon fluxes [e.g., shot noise, Johnson noise, quantization noise, thermal noise (if operation temperature $T<1K$), preamplifier noise, etc.] are all much less than the BPF\cite{jmp498}, in order to detect the PPFs generated by the HFGWs ($\nu\sim 10^{9}$ to $10^{12} Hz$, $h\sim10^{-21}$ to $10^{-23}$) in the braneworld\cite{cqgF33}, the requisite minimal accumulation time of the signals can be less or much less than $10^4 s$ [see  Appendix B]. Moreover, since the PPFs, Eqs. (\ref{eq67}), (\ref{eq69}), (\ref{eq75}), (\ref{eq80}), (\ref{eq88}) and the BPF, Eq. (\ref{eq64}) also have other very different  physical behaviors, such as the wave impedance, decay rate [the decay factor of the PPFs is $\exp(-\frac{r^2}{W^2})$, see Eqs. (\ref{eq67}), (\ref{eq69}), (\ref{eq80}) and (\ref{eq88}), while the decay factor of the BPF is $\exp(-\frac{2r^2}{W^2})$, see Eq. (\ref{eq64})], etc., then the displaying condition to the PPFs can be further relaxed. Besides, the ``rotation direction'' expressed by Fig.  \ref{fig9a}  is completely ``left-handed circular'' or completely  ``right-handed circular'', and the ``left-handed circular'' or  ``right-handed circular'' property depends on the phase factors in Eqs. (\ref{eq67}), (\ref{eq69}), (\ref{eq75}), (\ref{eq80}) and  (\ref{eq88}).\\
\indent In Fig.  \ref{fig9b}, the $n^{(1)}_{\phi}\propto\sin 2\phi$, and it includes following five cases: \\
(i) $n^{(1)}_{\phi-\oplus,l}$, Eq. (\ref{eq68}). This is the transverse PPF displaying the combination state of the  $\oplus$-type and the $l$-type polarizations (the tensor-mode and the scalar-mode gravitons) of the HFGWs. The PPF is from the EM response to the HFGWs in the coupling between the transverse static magnetic field  $\hat B_y^{(0)}$   and the GB in the 3DSR.\\
(ii) $n^{(1)}_{\phi-\otimes}$, Eqs. (\ref{eq70}) and (\ref{eq79}). They are the transverse PPFs displaying the pure $\otimes$-type polarization state (the tensor-mode gravitons) of the HFGWs. The PPFs are  from the EM response to the HFGWs in the coupling between the transverse static magnetic field $\hat B_x^{(0)}$ (or the transverse electric field $\hat E_y^{(0)}$) and the GB in the 3DSR, respectively.\\
(iii) $n^{(1)}_{\phi-y}$, Eq. (\ref{eq76}). This is the transverse PPF displaying the pure $y$-type polarization state (the vector-mode gravitons) of the HFGWs. The PPF is from the EM response to the HFGWs in the coupling between the longitudinal static magnetic field $\hat B_z^{(0)}$ and the GB in the 3DSR.\\
(iv) $n^{(1)}_{\phi-\oplus,l}$, Eq. (\ref{eq77}). This is the transverse PPF displaying the combination state of the $\oplus$-type and the $l$-type polarizations (the tensor-mode and the scalar-mode gravitons) of the HFGWs. The PPF is from the EM response to the HFGWs in the coupling between the transverse static electric field $\hat E^{(0)}_x$ and the GB in the 3DSR.\\
(v) $n^{(1)}_{\phi-x}$, Eq.  (\ref{eq87}). This is the transverse PPF displaying the  pure $x$-type polarization state (the vector-mode gravitons) of the HFGWs. The PPF is from the EM response to the HFGWs in the coupling between  the longitudinal static electric field $\hat E_y^{(0)}$   and the GB in the 3DSR.\\
\indent Unlike Fig.  \ref{fig9a}, here the ``rotation direction'' of the PPFs are not completely ``left-handed circular'' or not completely ``right-handed circular'', and it and the transverse BPF have the same angular distribution [see Eq. (\ref{eq64})]. Thus the displaying condition in Fig.  \ref{fig9b}  will be worse than that in Fig.  \ref{fig9a}. However, because the transverse PPFs in Fig.  \ref{fig9b}  and the BPF have other different physical behaviors, such as the different wave impedance, decay rate, and even  different propagating directions in the local region, it is always possible to display and distinguish the PPFs from the BPF.\\
\indent In Fig.  \ref{fig9c}, the $n^{(1)}_{\phi}\propto\sin\phi$, and it includes following three cases: \\
(i) $n^{(1)}_{\phi-x}$, Eq. (\ref{eq78c}). This is the transverse PPF displaying the  pure $x$-type polarization state (the vector-mode gravitons) of the HFGWs. The PPF is from the EM response to the HFGWs in the coupling between the transverse static electric field $\hat E_x^{(0)}$  and the GB in the 3DSR.\\
(ii) $n^{(1)}_{\phi-y}$, Eq. (\ref{eq81b}). This is the transverse PPF displaying the pure $y$-type polarization state (the vector-mode gravitons) of the HFGWs. The PPF is from the EM response to the HFGWs in the coupling between the transverse static electric field  $\hat E_y^{(0)}$  and the GB in the 3DSR.\\
(iii) $n^{(1)}_{\phi-b,l}$, Eq. (\ref{eq89b}). This is the transverse PPF displaying the combination state of the  $b$-type and the $l$-type polarizations (the  scalar-mode gravitons) of the HFGWs. The PPF is from the EM response to the HFGWs in the coupling between the longitudinal static electric field  $\hat E_z^{(0)}$   and the GB in the 3DSR.\\
\indent Unlike Fig.  \ref{fig9a}, here the ``rotation direction'' of the PPFs is not completely  ``left-handed circular'' or not completely  ``right-handed circular''. However, the best detection position of the PPFs is also the receiving surfaces at $\phi=\pi/2$ and $3\pi/2$ [see Fig.  \ref{fig9a}   and  \ref{fig9c}], where the PPFs have their peak values while the BPF vanishes.\\
\indent In Fig.  \ref{fig9d}, the $n^{(1)}_{\phi-\oplus,b,l}\propto\cos^2\phi$, Eq. (\ref{eq90}), and it is the transverse PPF  displaying the combination state of the  $\oplus$-type (the tensor-mode gravitons) and the $b$-type, $l$-type  polarizations (the  scalar-mode gravitons) of the HFGWs. The PPF is from the EM response to the HFGWs in the coupling between the transverse static magnetic field  $\hat B_x^{(0)}$   and the GB. It needs to be emphasized that the  $n^{(1)}_{\phi-\oplus,b,l}$, Eq. (\ref{eq69}), is also the transverse PPF displaying the combination state of the  $\oplus$-type,  the $b$-type and $l$-type  polarizations, but they have different angular distributions. The position of peak values of the  $n^{(1)}_{\phi-\oplus,b,l}$, Eq. (\ref{eq69}) are surfaces at $\phi=\pi/2$ and  $3\pi/2$, while the positions of peak values of the $n^{(1)}_{\phi-\oplus,b,l}$, Eq. (\ref{eq90}) are surfaces at $\phi=0, \pi$. Especially, the peak value areas of the PPF (the signal photon fluxes) are just the zero value area of the BPF (the background noise photon flux), Eq. (\ref{eq64}) and \mbox{Fig.  \ref{figure09}}. \\
\indent The PPFs, $n^{(1)}_{\phi-\oplus,b,l}$, Eq. (\ref{eq69}) and  Eq. (\ref{eq90}), are both from the same EM response to the HFGWs in the coupling between the transverse static magnetic field  $\hat B_x^{(0)}$ and the GB. This means that displaying the PPFs at such areas would have very strong complementarity. Moreover, like the PPFs expressed in Fig.  \ref{fig9a}, here, the PPF $n^{(1)}_{\phi-\oplus,b,l}$ is also completely ``left-handed circular'' or  completely  ``right-handed circular'' (see Fig.  \ref{fig9d}).\\
\indent In Fig.  \ref{fig9e}, the $n^{(1)}_{\phi-y}\propto\sin\phi\cos^2\phi$, Eqs. (\ref{eq78b}),  (\ref{eq78d}), (\ref{eq81a}) and (\ref{eq81c}), and it is the transverse PPF  displaying the pure $x$-type polarization (the vector-mode gravitons) of the HFGWs and  the pure $y$-type   polarizations (the  vector-mode gravitons) of the HFGWs, respectively. For the former, the PPF is from the EM response to the HFGWs in the coupling between the transverse static electric field $\hat E_x^{(0)}$ and the GB. For the latter, the PPF is from the EM response to the HFGWs in the coupling between the transverse static electric field  $\hat E_y^{(0)}$   and the GB. Clearly, the $n^{(1)}_{\phi-x}$ expressed by Eqs. (\ref{eq78b}), (\ref{eq78d}) and  the $n^{(1)}_{\phi-y}$ expressed by Eqs. (\ref{eq81a}), (\ref{eq81c}) are not completely ``left-handed circular'' or  completely  ``right-handed circular''. Besides, its angular distribution factor $\sin\phi\cos^2\phi=\frac{1}{2}\sin 2\phi\cos\phi$ is smaller than $\sin 2\phi$, Eq. (\ref{eq64}) of the BPF. Thus, distinguishability of the $n^{(1)}_{\phi-x}$ and $n^{(1)}_{\phi-y}$, in  Fig.  \ref{fig9e}  is worse than the PPFs in  Figs.  \ref{fig9a}  to     \ref{fig9d}. Nevertheless, considering obvious difference of other physical behaviors (e.g., the wave impedance, the decay rate, the propagating direction, etc.) between the PPFs in  Fig.  \ref{fig9e}  and the BPF in  Fig.  \ref{figure09}, their distinguishing is still possible.\\
\indent The above discussions show that the three polarization states (the  $\otimes$-type, the  $x$-type and the  $y$-type polarizations, i.e., the tensor-mode and the vector-mode gravitons) of the HFGWs can be clearly separated and distinguished.  On the other hand,  $\oplus$-type polarization (the tensor-mode gravitons), the  $b$-type and the  $l$-type polarizations (the scalar-mode gravitons) of the HFGWs are often expressed as their combination states to generate the PPFs. However, from the PPFs produced by these combination states, it is easy to calculate the PPFs  generated by the pure  $\oplus$-type, the pure $b$-type and the pure $l$-type polarizations, and thus we can completely determine these polarizations, respectively.\\
\indent From Eqs. (\ref{eq68}) to (\ref{eq69}) and  (\ref{eq89b}), we have
\begin{eqnarray}
	\label{eq91}
	&&n_{\phi-\oplus,b,l}^{(1)}=-n_{\phi-\oplus}^{(1)}+n_{\phi-b}^{(1)}+\sqrt{2}n_{\phi-l}^{(1)},\nonumber\\
	&&n_{\phi-\oplus,l}^{(1)}=n_{\phi-\oplus}^{(1)}+\frac{\sqrt{2}}{2}n_{\phi-l}^{(1)},\nonumber\\
	&&n_{\phi-b,l}^{(1)}=-n_{\phi-b}^{(1)}+\frac{\sqrt{2}}{2}n_{\phi-l}^{(1)},
\end{eqnarray}
where $n_{\phi-\oplus,b,l}^{(1)}$, Eq. (\ref{eq69}), $n_{\phi-\oplus,l}^{(1)}$, Eq. (\ref{eq68}), and $n_{\phi-b,l}^{(1)}$, Eq. (\ref{eq89b})  are the PPFs generated by the combination state of the $\oplus$-type, the $b$-type, the $l$-type polarizations, by the combination state of the $\oplus$-type, the $l$-type polarizations, and by the combination state of the $b$-type, the $l$-type polarizations, respectively.\\
\indent Clearly, $n_{\phi-\oplus}^{(1)}$,  $n_{\phi-b}^{(1)}$ and  $n_{\phi-l}^{(1)}$ in Eq. (\ref{eq91}) are the PPFs generated by the pure  $\oplus$-type, the pure $b$-type and the pure $l$-type polarizations of the HFGWs, respectively. By using  Eq. (\ref{eq91}), it is easy to calculate and find:
\begin{eqnarray}
	\label{eq92}
	n_{\phi-\oplus}^{(1)}=-\frac{1}{2}(n_{\phi-\oplus,b,l}^{(1)}-n_{\phi-\oplus,l}^{(1)}+n_{\phi-b,l}^{(1)}), 
\end{eqnarray}
\begin{eqnarray}
	\label{eq93}
	n_{\phi-b}^{(1)}=\frac{1}{2}(n_{\phi-\oplus,b,l}^{(1)}-n_{\phi-\oplus,l}^{(1)}-3n_{\phi-b,l}^{(1)}), 
\end{eqnarray}
\begin{eqnarray}
	\label{eq94}
	n_{\phi-l}^{(1)}=\frac{1}{\sqrt{2}}(n_{\phi-\oplus,b,l}^{(1)}-n_{\phi-\oplus,l}^{(1)}+n_{\phi-b,l}^{(1)}).
\end{eqnarray}
\indent Notice that the each term $n_{\phi-\oplus,b,l}^{(1)}$, $n_{\phi-\oplus,l}^{(1)}$ and $n_{\phi-b,l}^{(1)}$ of the right side in Eqs. (\ref{eq92}), (\ref{eq93}) and (\ref{eq94}) are directly measurable physical quantities. Therefore, the values and propagating directions of  $n_{\phi-\oplus}^{(1)}$, $n_{\phi-b}^{(1)}$ and $n_{\phi-l}^{(1)}$  can be completely confirmed.\\
\indent So far the PPFs produced by the six polarization states (the $\otimes$-type, the $x$-type, the $y$-type, the $\oplus$-type, the $b$-type and the $l$-type polarizations) of the HFGWs can be calculated and completely confirmed [e.g., see Eqs. (\ref{eq67}), (\ref{eq75}), (\ref{eq88}), (\ref{eq92}), (\ref{eq93}) and  (\ref{eq94}), respectively]. In other words, the six polarizations of the HFGWs can be clearly displayed and distinguished in the EM response of our 3DSR system.\\

\section{Numerical estimations of the perturbative photon fluxes}
\label{numerical}

\indent 1. Perturbative photon fluxes (signal EM waves) in the 3DSR system.\\
\indent Unlike the EM response to the HFGWs in the galactic-extragalactic background EM fields, the 3DSR is a closed cryogenic system with vacuum, which is shielded and isolated by superconductor materials from outside world. The EM response to the HFGWs in the 3DSR system has following important characteristics:\\
\indent (i) Since shielding of the 3DSR system to the EM fields from outside region, the large space-accumulation effect of the PPFs (the signal EM power fluxes) cannot enter and influence the EM fields inside the 3DSR system. However, the superconductor and the shell of the 3DSR system are transparent to the HFGWs. Thus, the PPFs inside the 3DSR system should be calculated by the HFGW amplitudes at the Earth (the far-field amplitudes, which are much less than the amplitudes in the near-field region of the HFGW source).\\
\indent (ii) Unlike the galactic-extragalactic background magnetic fields ($\hat B_x^{(0)}\sim10^{-9}$ to $10^{-11}T$), the background static magnetic fields $\hat B^{(0)}$ of the 3DSR system can reach up to $\sim10T$ or larger. The cooperation institute (High Magnetic Field Laboratory, Chinese Academy of Sciences) of our research team  has been fully equipped with the ability to construct the superconducting magnet\cite{CASmagnet}, and it is also the builder of the superconducting magnet for the Experimental Advanced Superconducting Tokamak (EAST) for  controlled nuclear fusion. The magnet can generate a static magnetic field with $\hat B^{(0)}=10T$ in an effective cross section with diameter of at least $80cm$ to $100cm$, and operation temperature can be reduced to $1K$ or less. Obviously, such magnetic field is much stronger than the galactic-extragalactic background magnetic fields, although typical spatial dimension of the former is only of the order of magnitude of a meter (typical laboratory dimension).\\
\indent (iii) The major noise sources of the EM signals from the case of galactic-extragalactic EM fields (see below), would be from the space microwave background, and the key noises of the case of laboratory based high magnet are from the microwave photons inside the 3DSR system, which are almost independent of the space background EM noise and the \mbox{cosmic} dusts. \\
\indent (iv) Because the PPFs are the first-order perturbations to the background EM fields and not the second-order perturbations, i.e., the PPFs (signal photon fluxes), Eqs. (\ref{eq67}) to (\ref{eq70}), (\ref{eq75}) to (\ref{eq81}) and (\ref{eq87}) to (\ref{eq90}), in the 3DSR system  are proportional to the amplitudes themselves ($h$) of the HFGWs and not their square $h^2$, e.g., see Eqs. (\ref{eq57}) and (\ref{eq60}) etc. (i.e., could be the PPFs in the galactic-extragalactic background magnetic fields). Then the parameter $h\hat B^{(0)}$ in the first-order PPFs [e.g. see Eqs. (\ref{eq67}) and (\ref{eq68})] will be much larger than the parameter $(h\hat B^{(0)})^2$ in the second-order PPFs [e.g. see Eqs. (\ref{eq35}) and (\ref{eq36})]. This property effectively compensates the weakness of the far-field amplitudes of the HFGWs in the 3DSR system. Of course, since the PPFs in the 3DSR system are always accompanied by the noise photons inside the system, which mainly are from the background photon fluxes caused by the GB. Thus, in order to identify the total signal photon flux at an effective receiving surface $\Delta s$, the time accumulation effect of the PPF must be larger than the effect of  the noise photon flux fluctuation at the receiving surface  $\Delta s$.\\
\indent As mentioned above, in order to display the relatively weak PPFs in the background noise photon fluxes (BPFs), we need an long enough  accumulation time of the signal [see   Appendix B]. For the typical parameters of the HFGWs predicted by the braneworld scenarios\cite{cqgF33,Andriot2017}, the observing  and distinguishing of the HFGWs would be quite possible due to their large amplitudes, higher frequencies and the discrete spectrum characteristics. The measurement of the relic HFGWs will face to big challenge, but it is not impossible (see Table \ref{tbb1}).\\

\begin{table*}[!htbp]
	\caption{\label{tbb1}%
		Displaying condition in 3DSR system (in laboratory scale) for the HFGWs (having possible additional polarizations) from some typical cosmological models and high-energy astrophysical process.}
	\begin{tabular}{cccccc}
		\hline
		Amplitude(A)  & Resonance& $n_{\phi(total)}^{(1)}(s^{-1}) $  &  $\Delta t_{min}(s)$  &Allowable upper 
		&Possible  verifiable  cosmological models   \\
		dimensionless &frequency&  &  & limit of noise photon
		& and astrophysical process \\
		&&&& flux $n^{(0)}_{\phi(total)}(s^{-1})$&  \\
		\hline
		$10^{-23}$&$3\times10^9Hz$&$\sim10^{9}$&$\sim10^4$&$\sim10^{22}$&Braneworld\cite{cqgF33}\\
		$10^{-25}$&$3\times10^9Hz$&$\sim10^{7}$&$\sim10^4$&$\sim10^{18}$&Braneworld\cite{cqgF33}\\
		$10^{-26}$&$3\times10^8Hz$&$\sim10^{6}$&$\sim10^5$&$\sim10^{18}$&Short-term anisotropic inflation\cite{Ito.anisotropic.2016}\\
		$10^{-27}$&$3\times10^9Hz$&$\sim10^{5}$&$\sim10^5$&$\sim10^{16}$&Interaction of astrophysical plasma\\
		&&&&& with intense EM radiation\cite{prd044017}\\
		$10^{-29}$&$3\times10^9Hz$&$\sim10^{3}$&$\sim10^6$&$\sim10^{12}$&Pre-big-bang\cite{sciam290,pr373}\\
		$10^{-30}$&$3\times10^9Hz$&$\sim10^{2}$&$\sim10^6$&$\sim10^{10}$&Quintessential inflationary\cite{prd123511,cqg045004,Giovannini2014} or \\
		&&&&&  upper limit of ordinary inflationary\cite{0504018,084022} \\
		\hline
	\end{tabular}
\end{table*}
\indent Table \ref{tbb1} shows the displaying conditions of the HFGWs for some typical cosmological models and high-energy astrophysical processes,
where $n_{\phi(total)}^{(1)}$ is the total signal photon flux at the receiving surface $\Delta s$ ($\Delta s\sim 3\times 10^{-2} m^2 $ ), and $n_{total}^{(0)}$ is the allowable upper limit of the total noise photon flux at the surface
$\Delta s$ for various values of the HFGW amplitudes, and the $\Delta t_{min}$  [see Eq.(\ref{eqb1}) in Appendix B] is the requisite minimal accumulation time of the signals;  $\nu_e=\nu_g=3\times10^9Hz$ or  $3\times10^8Hz$ (the resonance frequency); the background
static magnetic field $B^{(0)}$ is 10T; the interaction dimension $\Delta \textit{z}$ is 60cm (typical dimension of the static magnetic field in our 3DSR); the power of the Gaussian beam is $\sim10W$
and the operation temperature should be less than 1K.  \\
\indent Notice that for the GB of $P\sim10W$ in the 3DSR, the maximum of $n^{(0)}_{\phi(total)}$ at the receiving surface $\Delta s=3\times10^{-2}m^2$ is about $\sim10^{22}s^{-1}$. This means that even if the peak values of the noise photon flux and the PPFs appear at the same receiving surface (e.g., see Fig. \ref{fig9b}), the $\Delta t_{min}$ can be limited in $\sim10^4s$ (e.g. the first case in   Table \ref{tbb1}). In fact, in many cases discussed in this paper, the peak position of such two kinds of photon fluxes do not appear at the same receiving surface, and especially, the peak value positions (e.g., see Figs. \ref{fig9a}, \ref{fig9c} and \ref{fig9d}) of the signal photon fluxes are just the zero value areas ($\phi=0$, $\pi/2$, $\pi$, $3\pi/2$) of the background noises photon flux  $n^{(0)}_{\phi}$ (see Fig.  \ref{figure09}). In this case the displaying condition can be further relaxed.\\

\indent In Table \ref{tbb1}, the $\nu\sim 3\times10^{8}$ to $3\times10^{9} Hz$  is a typical peak value region of the spectrum of the HFGWs in the braneworld scenarios\cite{cqgF33}, the short-term anisotropic inflationary model \cite{Ito.anisotropic.2016}, the pre-big-bang model\cite{sciam290,pr373} and the quintessential inflationary model\cite{prd123511,cqg045004,Giovannini2014}.\\
\indent It is very interesting to compare the displaying and  distinguishing conditions for the case in the galactic- extragalactic fields (e.g. see part 2 of this section and section \ref{perturbativeEMWs})  and in the 3DSR (section \ref{3DSR}). In section \ref{perturbativeEMWs}, the large space accumulation effect caused by the galactic-extragalactic fields is discussed, and the section \ref{3DSR} shows the longer time accumulation effect in the strong background static magnetic field and the resonance response of the GB in the 3DSR. They all can effectively compensate the weakness of the HFGW amplitudes. Clearly, they would be highly complementary to each other. Moreover, the PPFs produced in the 3DSR system by the relic HFGWs in the quintessential inflationary, the pre-big-bang and the short-term anisotropic inflationary models, would only be  $10^2s^{-1}$ to  $10^6s^{-1}$ (see Table \ref{tbb1}), which are much less than the maximum of the transverse BPF (noise photon flux). However, using very different physical behaviors between the PPFs and the BPFs, especially their very different strength distributions and propagating directions [e.g., see Figs.  \ref{figure09},  \ref{fig9a},  \ref{fig9b},   \ref{fig9d}, and Eqs. (\ref{eqb3}) and (\ref{eqb4}) in Appendix B], it is still possible to reduce the BPF to the allowable upper limits at a suitable receiving surface (see Table \ref{tbb1}). Considering their very different wave impedances (four orders of magnitude at least\cite{LiNPB2016}), the displaying of the transverse PPFs produced by the relic HFGWs may also be possible though face to big challenge.\\

\begin{table*}[!htbp]
	\caption{\label{table02}%
		The EM response to the KK-HFGWs from the braneworld \cite{cqgF33} in the galactic-extragalactic magnetic fields, where $N^{(2)}$ is the perturbative photon flux densities ($m^{-2}$), $P$ is corresponding EM signal power flux densities ($Wm^{-2}$) at the Earth, and here we assume that the near-field amplitudes $a_{\oplus}$, $a_{\otimes}$, $a_{x}$, $a_{y}$, $a_{b}$ and $a_{l}$ of the HFGWs in Eqs. (\ref{eq95}) to (\ref{eq99}), have the same strengths, then the PPFs calculated by such equations have approximately the same order of magnitude. Here, ``$a$'' represents typical near-field amplitude of the  $a_{\oplus}$, $a_{\otimes}$, $a_{x}$, $a_{y}$, $a_{b}$, $a_{l}$.
	}
	\begin{tabular}{ccc}
		\hline
		Background magnetic field $\hat B^{(0)}$ & $\nu_g\sim10^9Hz$  & $\nu_g\sim10^{12}Hz$  \\
		\hline
		$\hat B^{(0)}=10^{-10}T$ &    &   \\	
		
		$a\sim10^{-12}$ &$N^{(2)}=6\times10^{4}m^{-2} (P^{ }=10^{-19}Wm^{-2})$&$N^{(2)}=6\times10^{7}m^{-2} (P^{ }=10^{-13}Wm^{-2})$\\
		$a\sim10^{-10}$ &$N^{(2)}=6\times10^{8}m^{-2} (P^{ }=10^{-15}Wm^{-2})$&$N^{(2)}=6\times10^{11}m^{-2} (P^{ }=10^{-9}Wm^{-2})$\\
		$a\sim10^{-8}$&$N^{(2)}=6\times10^{12}m^{-2} (P^{ }=10^{-11}Wm^{-2})$&$N^{(2)}=6\times10^{15}m^{-2} (P^{ }=10^{-5}Wm^{-2})$\\
		\hline
		$\hat B^{(0)}=10^{-11}T$ &    &   \\	
		
		$a\sim10^{-12}$ &$N^{(2)}=6\times10^{2}m^{-2} (P^{ }=10^{-21}Wm^{-2})$&$N^{(2)}=6\times10^{5}m^{-2} (P^{ }=10^{-15}Wm^{-2})$\\
		$a\sim10^{-10}$ &$N^{(2)}=6\times10^{6}m^{-2} (P^{ }=10^{-17}Wm^{-2})$&$N^{(2)}=6\times10^{9}m^{-2} (P^{ }=10^{-11}Wm^{-2})$\\
		$a\sim10^{-8}$&$N^{(2)}=6\times10^{10}m^{-2} (P^{ }=10^{-13}Wm^{-2})$&$N^{(2)}=6\times10^{13}m^{-2} (P^{ }=10^{-7}Wm^{-2})$\\
		\hline
	\end{tabular}
\end{table*}
2. The perturbative photon fluxes (PPFs) in the EM response of galactic-extragalactic background stable EM fields to the HFGWs.\\
\indent Because there are very widely distributed galactic- extragalactic stable or quasi-stable stable EM fields (e.g. especially, background magnetic fields\cite{galacticB.RevModPhys.74.775}), the interaction of the HFGWs with the galactic-extragalactic stable magnetic fields could generate the PPFs when the HFGWs passing through such background magnetic fields.\\
\indent In the above discussion, we assume that the GWs are planar waves. For the observers and detection systems in the far field region from the GW sources,  such assumption, is obviously reasonable. However, if we study a large space accumulation effect in the EM response to the HFGWs, which is in large distance from their sources to the Earth, then such HFGWs should be considered as the spherical GWs for GW sources in the local regions, i.e., these HFGW sources would be in approximately point-like distribution. This means that the amplitudes of the HFGWs emitted by such sources would be inversely proportional to the propagating distance $z$ (i.e., along the $z$-axis in our coordinate system). Then the amplitudes $A_{\oplus}$, $A_{\otimes}$, $A_{x}$, $A_{y}$, $A_{b}$ and $A_{l}$ of the HFGWs in the Eqs. (\ref{eq35}), (\ref{eq36}), (\ref{eq39}), (\ref{eq40}), (\ref{eq45}), (\ref{eq46}), (\ref{eq51}), (\ref{eq52}) and (\ref{eq57}) to (\ref{eq60}) should be replaced by  $a_{\oplus}z_0/z$, $a_{\otimes}z_0/z$, $a_{x}z_0/z$, $a_{y}z_0/z$, $a_{b}z_0/z$ and $a_{l}z_0/z$, respectively, where $z_0$ is the reasonable reference distance of the near-field region, which is much less than the distance between the Earth and the HFGW sources, and the  $a_{\oplus}$, $a_{\otimes}$, $a_{x}$, $a_{y}$, $a_{b}$ and $a_{l}$ are the amplitudes in the near-field region of the HFGW sources. Obviously,  $a_{\oplus}$, $a_{\otimes}$, $a_{x}$, $a_{y}$, $a_{b}$ and $a_{l}$ are much larger than the amplitudes $A_{\oplus}$, $A_{\otimes}$, $A_{x}$, $A_{y}$, $A_{b}$ and $A_{l}$  in the far-field region.\\
\indent On the other hand, all of the perturbative EM fields are proportional to the propagating distance $z$ due to the space accumulation effect caused by the same or almost the same propagating velocity between the perturbative EM waves (PPFs) and the HFGWs (gravitons)  [also, see Eqs. (\ref{eq35}), (\ref{eq36}), (\ref{eq39}), (\ref{eq40}), (\ref{eq45}), (\ref{eq46}), (\ref{eq51}), (\ref{eq52}) and (\ref{eq57}) to (\ref{eq60})]. Then the parameter $z$ in the numerator and in the denominator in these equations will be canceled each other. In other words, the strength of the perturbative EM fields (i.e., the PPFs) would be a composite effect of the space accumulation of the EM signals and the decay of the HFGWs. According to previous treatment method \cite{PRD104025} of such effect   and the above equations, in the same way,  the PPFs generated by the EM response in the background stable magnetic fields, can be reduced to following forms, respectively,
\begin{eqnarray}
	\label{eq95}
	n_{\oplus,l}^{(2)}=\frac{k_g^2z_0^2c}{8\mu_0\hbar\omega_e}[(a_{\oplus}+\frac{\sqrt{2}}{2}a_l) \hat B_y^{(0)}]^2, 
\end{eqnarray}
\begin{eqnarray}
	\label{eq96}
	n_{\otimes}^{(2)}=\frac{k_g^2z_0^2c}{8\mu_0\hbar\omega_e}[(a_{\otimes}\hat B_x^{(0)})^2+(a_{\otimes}\hat B_y^{(0)})^2], 
\end{eqnarray}
\begin{eqnarray}
	\label{eq97}
	n_{\oplus,b,l}^{(2)}=\frac{k_g^2z_0^2c}{8\mu_0\hbar\omega_e}[(a_{b}-a_{\oplus}+\sqrt{2}a_l) \hat B_x^{(0)}]^2, 
\end{eqnarray}
\begin{eqnarray}
	\label{eq98}
	n_{x}^{(2)}=\frac{k_g^2z_0^2c}{8\mu_0\hbar\omega_e}(a_x\hat B_z^{(0)})^2, 
\end{eqnarray}
\begin{eqnarray}
	\label{eq99}
	n_{y}^{(2)}=\frac{k_g^2z_0^2c}{8\mu_0\hbar\omega_e}(a_y\hat B_z^{(0)})^2, 
\end{eqnarray}
\indent where $z_0$ is the reference distance in the near-field region. If such HFGWs are from point-like objects orbiting a braneworld black hole, then $z_0$ is at least in the scale of the horizon size of the black hole, and $z_0\approx10^4m$ due to the horizon size of a black hole with a mass of $10M_{\odot}$\cite{cqgF33,PRD104025}.\\
\indent Notice that $a_{\oplus}$, $a_{\otimes}$, $a_{x}$, $a_{y}$, $a_{b}$ and $a_{l}$ in Eqs.  (\ref{eq95}) to (\ref{eq99}) are the amplitudes of the HFGWs in the near-field region of the HFGW source, which are much larger than the amplitudes  $A_{\oplus}$, $A_{\otimes}$, $A_{x}$, $A_{y}$, $A_{b}$ and $A_{l}$ in the far-field region (e.g., the Earth) of the HFGW source. As mentioned above, this is because the large space accumulation effect would compensate the decay of the spherical HFGWs and their weakness of the far-field amplitudes.\\
\indent Fortunately, related observation shows that the distribution region of the stable background magnetic field (see Fig. \ref{figure08}) in our galaxy is nearly few thousand light-year distance ($\sim10^{19}m$\cite{galacticB.RevModPhys.74.775,0004-637X-663-1-258}), and such magnetic field basically keeps fixed direction  and intensity $\sim10^{-9}$ to $10^{-10}$Tesla, thus this background magnetic field would provide an effective space-accumulation effect to the PPFs produced by the HFGWs.\\
\indent According to related estimation for the distance of possible HFGW sources of the braneworld scenarios [e.g., see Ref. \cite{cqgF33}]  in our galaxy, they may be $\sim10^{18}m$ to $\sim10^{19}m$ away from the Earth. In this case, if the amplitudes of the HFGWs ($\nu\sim 10^{9}$ to $10^{14}Hz$) can reach up to the predicted values once they arrive at the Earth, i.e., $h\sim10^{-25}$ (lower bound) to  $h\sim10^{-21}$ (upper bound), then the near-field amplitude of the HFGWs would be  $h_0\sim10^{-6}$ to $10^{-10}$ due to the horizon size of the black hole with a mass of $10M_{\odot}$.\\
\indent Table \ref{table02} shows the estimations of the PPFs (the signal EM power flux densities) generated by the EM response of the background stable magnetic fields to the HFGWs. Here $\hat B^{(0)}\sim10^{-11}T$ to $\sim10^{-10}T$; the propagating distance $\Delta z\sim10^{19}m$ of the HFGWs from the source to the Earth; the amplitudes of the HFGWs at the Earth are $h\sim10^{-21}$ to $10^{-25}$ \cite{cqgF33}; then the near-field amplitudes of such HFGWs would be $a\sim10^{-6}$ to $10^{-10}$, roughly.\\
\indent Obviously, the PPFs (the EM signals) in Table \ref{table02} are larger than the minimal detectable EM power $\sim10^{-22}Wm^{-2}$ to $\sim10^{-24}Wm^{-2}$ under current technological condition [e.g., the Five-hundred-meter Aperture Spherical Telescope (FAST)\cite{FAST2011,liDiRDS20375}, which has been constructed in 2017 in Guizhou province of China. However, above EM signals and the space background EM noise are often accompanied together. Thus distinguishing of the EM signals from the background EM noise at the Earth will be a major challenge. Fortunately,  the PPFs (the EM signals) produced by the HFGWs from the braneworld   also contain important characteristics of the HFGWs, such as the discrete spectrum property, special waveform, strength distribution and very-high frequency features. Thus, it is always possible to distinguish and display the PPFs (the EM signals) from the background EM noise.\\
\indent It should be pointed out again that if the propagating direction of the GWs (including the HFGWs) and the pointing direction of the background EM fields are parallel to each other (i.e., only the longitudinal EM fields $\hat E_z^{(0)}$, $\hat B_z^{(0)}$ exist here), then GWs in the GR framework do not generate any perturbative effect to the EM fields\cite{nuovo129,prd2915}. Unlike the physical behavior of the GWs in the GR framework, the GWs having additional polarization states will generate perturbative EM fields to the background EM fields in any case, even if the pointing direction of the background EM fields and the propagating direction of the HFGWs are parallel to each other [see Eqs. (\ref{eq57}), (\ref{eq58}), (\ref{eq98}), (\ref{eq99})], which are just the perturbation effects produced by the pure additional polarization states (the $x$-type and the $y$-type polarizations, i.e., the vector mode gravitons) of the HFGWs.\\
\indent In this case, displaying condition of the HFGWs can be greatly relaxed. In other words, this is an important symbol to display and distinguish the HFGWs in the GR framework and the HFGWs beyond the GR, whatever  in the astrophysical scale  (e.g. see section \ref{perturbativeEMWs}) or  in the laboratory dimension \mbox{ (e.g. in 3DSR, section  \ref{3DSR})}.
\\

\section{Concluding remarks}
\label{conclusion}

\indent In this paper,   for the first time we address the EM counterparts of the HFGWs having additional polarization states in the EM response, and it shows the   \mbox{separability} and    detectability  of the possible six polarization states of the HFGWs in the EM response.

\indent 1. For the EM response to  the HFGWs in the 3DSR system and for the EM response  to  the HFGWs in the galactic- extragalactic background stable EM fields, the pure  $\otimes$-type polarization (the tensor-mode gravitons), the pure  $x$-type and   the pure  $y$-type  polarization (the vector-mode gravitons) of the HFGWs can independently generate the PPFs. The    $\oplus$-type polarization (the tensor-mode gravitons),  the  $b$-type  and the  $l$-type  polarization (the scalar-mode gravitons) in the EM response produce the PPFs in their different combination states. However,   the EM perturbations generated by the pure    $\oplus$-type, the pure    $b$-type and the pure    $l$-type polarizations  can be directly calculated and can be completely determined  from such combination states, respectively. Therefore, the  six polarization states of the HFGWs all have the separability and detectability.\\
\indent 2. The coupling between the background longitudinal static EM fields  $\hat E_z^{(0)}$, $\hat B_z^{(0)}$ and the GB in the 3DSR would only produce the EM response to the $x$-type, the $y$-type polarizations and the combination state of the   $b$-type and   the $l$-type polarizations [see Eqs. (\ref{eq75}), (\ref{eq76}), (\ref{eq87}), (\ref{eq88})  and (\ref{eq89a}) to (\ref{eq89c})]. 
The interaction of the HFGWs with the galactic- extragalactic longitudinal stable EM fields $\hat E_z^{(0)}$, $\hat B_z^{(0)}$ would only generate the EM response to the $x$-type and the $y$-type polarizations of the HFGWs [see Eqs. (\ref{eq57}) and (\ref{eq58})]. 
In other words, the longitudinal background static EM fields only produce EM response to the additional polarizations (the $x$-type, the $y$-type, the $b$-type and the $l$-type polarization states) of the HFGWs, and do  not generate any EM response to the traditional  $\otimes$-type or  $\oplus$-type polarizations in the GR framework. This would provide an effective way to display the pure additional polarization states of the HFGWs.\\
\indent 3. The PPFs [the EM perturbative power fluxes, see Eqs. (\ref{eq95}) to (\ref{eq99})] generated by the HFGWs having additional polarization states are obviously larger than the PPFs  generated by the HFGWs in framework of the GR [e.g., when $a_x$, $a_y$, $a_b$ and $a_z$ equal to zero in the Eqs. (\ref{eq95}) to (\ref{eq99})]. This means that the former would cause larger GW radiation damping than the latter for the local GW sources surrounded by stable magnetic fields (e.g. the galactic-extragalactic magnetic fields). Clearly, the above effect is not only valid to the HFGWs, but also would be still valid to the intermediate-frequency and the low-frequency GWs generated by the neutron star binaries and by the black hole mergers in the galactic-extragalactic magnetic fields. Thus accurate observation of their period evolutions and related intermediate-frequency and the low-frequency EM radiations (i.e., related EM counterparts) will be significant works.\\
\indent 4. In a specific transverse direction, i.e., the x-direction in the 3DSR system discussed in this paper, the BPF  $n_x^{(0)}=0$. However, the PPFs $n_x^{(1)}$ propagating along the $x$-direction have significant non-vanishing values [see Eqs. (\ref{eqb3}) to (\ref{eqb5}) in Appendix B]. In this case, the displaying condition in Table \ref{tbb1}  can be greatly relaxed. In other words, the PPFs produced by the HFGWs from the braneworld\cite{cqgF33,Andriot2017}, by HFGWs predicted by the short-term anisotropic inflationary model\cite{Ito.anisotropic.2016} and by HFGWs from the interaction\cite{prd044017} of astrophysical plasma with intense EM waves, can almost be instantaneously  displayed;  the requisite minimal accumulation time of the signals for displaying relic  HFGWs expected by the pre-big-bang\cite{pr373,sciam290} and by the quintessential inflationary\cite{prd123511, cqg045004,Giovannini2014} models, would be further relaxed effectively. Thus, utilizing a highly orientational receiving surface to display the PPFs, will be very useful.\\
\indent 5.  The longitudinal perturbative electric field $\tilde E_z^{(1)}$ plays an important role in the 3DSR system, i.e., the PPFs [see Eqs. (\ref{eq78b}) to (\ref{eq78d}) and Eqs. (\ref{eq89a}) to (\ref{eq89c})] produced by the  resonance effect between the GB and the longitudinal perturbative electrical field $\tilde E_z^{(1)}$ can effectively display and distinguish the additional polarization states (the $x$-type,   the $b$-type and the $l$-type polarizations) of the HFGWs.
Whereas, in the EM response of the galactic-extragalactic background stable electric field $\hat E^{(0)}$ to the HFGWs, the average values of the transverse PPFs equal to zero [see Eq.(\ref{eq61})] due to the phase difference of $\pi/2$ between the longitudinal perturbative electric field,  Eq. (\ref{eq31}), and the transverse perturbative magnetic fields,  Eqs. (\ref{eq28}) and (\ref{eq30}), i.e., the longitudinal perturbative electric field $\tilde E_z^{(1)}$ does not provide any contribution for the PPFs in such case.  \\
\indent 6. In the EM response to the HFGWs in the 3DSR system, the number of the PPFs [e.g., see Eqs. (\ref{eq67}), (\ref{eq68})] is independent of the frequency due to $k_g=\omega_g/c$ [i.e., the related strength of the signal EM power fluxes are proportional to the frequencies of the HFGWs]. 
In the EM response to the HFGWs in galactic- extragalactic stable magnetic fields, the strengths of the PPFs [e.g., see Eqs. (\ref{eq35}), (\ref{eq36})] are not only proportional to the square of amplitudes of the HFGWs, but also proportional to their frequencies due to $k_g=\omega_g/c$ [i.e., the related strengths of the signal EM power fluxes are proportional to the square of their frequencies].  
Therefore, very high frequency can effectively compensate the weak HFGW amplitudes. Especially for the HFGWs predicted by the braneworld scenarios\cite{cqgF33}, since the amplitudes of such HFGWs almost have the same order of magnitude in the frequency band of $\sim 10^{9} Hz$ to $  10^{12} Hz$, and because of $\hbar\omega_g=KT$ (K is the Boltzmann constant), the power of the PPFs of $10^{12} Hz$ will be roughly $10^3$ times  larger than that of the PPFs of  $10^{9} Hz$, and requirement of operating temperature can be relaxed to $140K$ from $0.14K$. In this case, for the background photon flux $n^{(0)}_{\phi}$ having the same power, the number of the  $n^{(0)}_{\phi}$ of $10^{12} Hz$ will be reduced  $10^3$ times for the  $n^{(0)}_{\phi}$ of $10^{9} Hz$. Then the requisite minimal accumulation time of the signals will be reduced by almost one thousand times [see Eq. (\ref{eqb1})].\\

\indent Besides, we need to mention two points:\\
\indent (1) Since the 3DSR is a detection system fixed on the Earth, so the 3DSR will have a rotation period of 24 hours to any GW sources in the local space region due to the rotation of the Earth. Thus the background ``transverse'' and ``longitudinal'' EM fields in the 3DSR will be periodically changed with the rotation of the Earth. In fact, this is an issues on the relationship between the characteristic parameters (i.e., each polarization component) of the HFGWs  and the coordinate rotation. In our 3DSR system, this is an issue on the relationship between the EM response to the HFGWs and the coordinate rotation. Nevertheless, the 3DSR has a special and definite direction, namely, the positive direction of the symmetrical axis of the GB (i.e. the $z$ direction in section \ref{3DSR}). We have showed\cite{FYLi_EPJC_2008} that only when the propagating direction of the HFGWs and the positive direction of the symmetrical axis of the GB are the same, the PPFs  reach their maximum. If they are perpendicular or opposite to each other, then the PPFs will be one or two orders of magnitude smaller than their peak values. Thus we can determine the propagating direction of the HFGWs by the peak moments of PPFs. The preliminary calculations show that within about two hours around the peak moments, the PPFs can basically maintain the intensity at the same orders of magnitude as the peak values. This means that we can maintain an effective signal accumulation time of $\sim7\times10^3s$ for the PPFs, i.e., it basically approaches $10^4s$ in Table  \ref{tbb1}. This is satisfactory. Moreover, the 3DSR is an EM detection system in small scale (order of magnitude of a meter). Thus its spatial orientation can also be effectively adjusted to the best or near optimal direction for the possible HFGW sources.\\
\indent (2) The detection frequency band of the 3DSR is $\sim10^8Hz$ to $10^{12}Hz$ or higher, and the observational frequencies (for EM signals) of the FAST is $\sim7\times10^7Hz$ to $3\times10^9Hz$\cite{li_nan_pan_2012,liDiRDS20375}. This means that the detection frequency bands of the 3DSR and   the FAST are overlapping partly in the GHz band. Therefore, the cooperation and coincidence experiments of them will have very strong complementarity  to distinguish and display the possible all  six polarizations of the HFGWs in microwave frequency band. \\
\indent The two points mentioned above will be discussed and studied in detail elsewhere.\\

~\\
~\\
~\\
~\\

\appendix
\section{A new group of special solutions of the Gaussian-type photon flux for the Helmholtz equation}
\label{appendixA}
\indent In this paper, the 3DSR is actually a coupling system between the background static EM fields and the Gaussian-type photon flux (Gaussian beam). Under the condition of the resonance response to the HFGWs, the PPFs (the signal photon fluxes) and the BPFs (including other noise photons) have very different physical behaviors in the local regions. Thus it makes the 3DSR system having very low standard quantum limit. In Refs.\cite{prd104008,FYLi_EPJC_2008} we have given several coupling forms between the GB and the background static magnetic field. Here we select a new group of solutions of the Helmholtz equation, and give their complete expressions. Moreover, in order to display and distinguish effectively all six polarization states of the HFGWs, here such GB does not only couple with the transverse static EM fields, but also the longitudinal static EM fields. In this case, it is possible to display and distinguish  all of the different polarization states of the HFGWs.\\
\indent General form of the circular mode GB of fundamental frequency is\cite{Yariv}
\begin{eqnarray}
	\label{eqa1}
	&&	\psi=\frac{\psi_0}{\sqrt{1+(z/f)^2}}\exp(\frac{-r^2}{W^2})\exp\{i[(k_e z-\omega _e t)\nonumber\\
	&&	-tan^{-1}\frac{z}{f}+\frac{k_er^2}{2R}+\delta]\},
\end{eqnarray}
where $\psi_0$ is the amplitude of electrical field of the GB, $f=\pi W_0^2/\lambda_e$, $W=W_0\sqrt{1+(z/f)^2}$, $R=z+f^2/z$. The $W_0$ is the minimum spot radius, $R$ is the curvature radius of the wave front of the GB at $z$, $\omega_e$ is the angular frequency, $\lambda_e$ is the EM wavelength, the $z$-axis is the symmetrical axis of the GB, and $\delta$ is a phase factor.\\
\indent According to Eq. (\ref{eqa1}) and using the   the non-divergence condition $\nabla\cdot\bf{\tilde{E}}^{(0)}=0$ in the free space,  and 
$\rm{\bf\tilde{B}^{(0)}}=-i/\omega_e \cdot\bigtriangledown\times \rm{\bf\tilde{E}^{(0)}}$, a group of special solutions we selected are as follows, and the components of the GB in the Cartesian coordinates can be given by
\begin{eqnarray}
	\label{eqa2}
	&&\tilde E_x^{(0)}=\psi_{ex}=\psi=\frac{\psi_0}{\sqrt{1+(z/f)^2}}\nonumber
\end{eqnarray}
\begin{eqnarray}
	&&\exp(\frac{-r^2}{W^2})\exp\{i[(k_e z-\omega _e t)-tan^{-1}\frac{z}{f}+\frac{k_er^2}{2R}+\delta ]\},\nonumber
\end{eqnarray}
\begin{eqnarray}
	&&\tilde E_y^{(0)}=\psi_{ey}=0,\nonumber\\
	&&\tilde E_z^{(0)}=\psi_{ez}=2xF_1({\bf{x}},k_e,W)=2rcos\phi F_1({\bf{x}},k_e,W).~~~~~~~~
\end{eqnarray}
\begin{eqnarray}
	\label{eqa3}
	&&\tilde B_x^{(0)}=\psi_{bx}=-\frac{i}{\omega_e}\frac{\partial \psi_{ez}}{\partial y}=\frac{\sin 2\phi}{\omega_e}F_2(x,k_e,w),\nonumber
\end{eqnarray}
\begin{eqnarray}
	&&\tilde B_y^{(0)}=\psi_{by}=-\frac{i}{\omega_e}(\frac{\partial \psi_{ex}}{\partial z  }-\frac{\partial \psi_{ez}   }{\partial x  })=\frac{1}{\omega_e}[F_4({\bf{x}},k_e,W)\nonumber\\
	&&-i(2F_1({\bf{x}},k_e,W)+\cos^2\phi F_3({\bf{x}},k_e,W))],\nonumber\\
	&&\tilde B_z^{(0)}=\psi_{bz}=\frac{i}{\omega_e}\frac{\partial \psi_{ex}}{\partial y}=\frac{-\sin \phi}{\omega_e}[\frac{k_er}{R}+\frac{i2r}{W_0^2[1+(z/f)^2]}]\psi.\nonumber\\
\end{eqnarray}
\begin{eqnarray}
	\text{where,}~~~	\label{eqa4}
	F_1=\int(\frac{1}{W^2}-i\frac{k_e}{2R})\psi dz,
\end{eqnarray}
\begin{eqnarray}
	\label{eqa5}
	&&	F_2=\int\{\frac{3k_er}{W_0^2[1+(z/f)^2]R}\nonumber\\
	&&	+i[\frac{k_e^2r}{2R}+\frac{2r}{W_0^4[1+(z/f)^2]}]\}\psi dz,  ~~~~~~
\end{eqnarray}
\begin{eqnarray}
	\label{eqa6}
	F_3=\int(\frac{k_e^2r}{2R}-\frac{2r^2}{W^4}+i\frac{6k_er^2}{W^2R})\psi dz,
\end{eqnarray}
\begin{eqnarray}
	\label{eqa7}
	&&F_4=\{k_e-\frac{1}{f+z^2/f}-\frac{k_er^2[1-(f/z)^2]}{2R^2}\nonumber\\
	&& +i[\frac{z}{f^2[1+(z/f)^2]}+ \frac{2r^2}{W^2R}]  \}\psi.
\end{eqnarray}	
\indent According to Eqs. (\ref{eqa1}) to (\ref{eqa7}), we obtained the strength distribution (see Fig. \ref{figure09}) of the transverse background photon flux $n_{\phi}^{(0)}$ in the cylindrical polar coordinates. As demonstrated in this paper, the PPFs (signal photon fluxes) and the BPF (the dominated noise photon flux)   have very different physical behaviors (such as the strength distribution, propagating direction, decay rate, wave impedance etc.) in the local regions. This is one  important physical mechanism of the displayability and separability between the PPFs and the BPF, including the different polarizations of the HFGWs.\\

\section{Displaying conditon in the background noise photon flux fluctuation.}
\label{appendixB}
\begin{figure}[!htbp]
	\centerline{\includegraphics[scale=0.65]{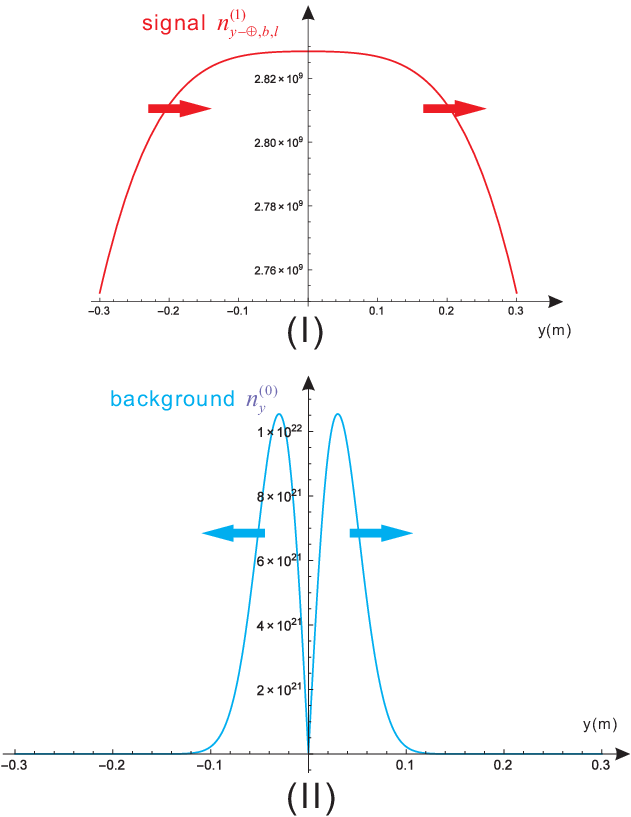}}
	\caption{\footnotesize{Comparison of diagrams of signal photon flux $n_{y-\oplus,b,l}^{(1)}$ (I) and background photon flux $n_{y}^{(0)}$ (II). It's shown that the $n_{y-\oplus,b,l}^{(1)}$ propagates always in the positive direction of y-axis, and the background noise photon flux $n_{y}^{(0)}$ propagates in the opposite directions of the y-axis of  $y>0$ and $y<0$.  }}
	\label{figureb1}
\end{figure}

\indent Unlike the B-mode experiment in CMB for the very low frequency relic GWs (where major noise is from the cosmic dust), here the noise sources would be the microwave noise photons inside the 3DSR, and they are almost independent of the cosmic dust. These noise photons include the BPF caused by the GB, and other noise photon fluxes (shot noise, Johnson noise, quantization noise, thermal noise, preamplifier noise, diffraction noise, etc). However, the above other noise photon fluxes are all much less than the BPF if operation temperature $T<1K$ for the thermal noise\cite{jmp498}. Thus, although the signal photon fluxes are always accompanied by such noise photons, our attention can mainly focus on the BPF fluctuation caused by the GB. In other words, once the PPFs can be displayed in the BPF fluctuation, the influence of the all other noise photon flux fluctuation can be neglected.\\
\indent The displaying condition in the noise photon flux fluctuation can be given by
\begin{eqnarray}
	\label{eqb1}
	n_{\phi(total)}^{(1)}\Delta t~\geqslant~\sqrt{n_{\phi(total)}^{(0)}\Delta t},~
	then  ~\Delta t~\nonumber\\
	\geqslant~n_{\phi(total)}^{(0)}/{[n_{\phi(total)}^{(1)}]}^2 = \Delta t_{min},
\end{eqnarray}\mbox{}
where $\Delta t_{min}$ is the requisite minimal accumulation time of the signals, and
\begin{equation}
	\label{eqb2}
	n_{\phi(total)}^{(1)}= \int\limits_{~~\Delta s} n_{\phi}^{(1)}ds,
	~~\text{and}~~n_{\phi(total)}^{(0)}= \int\limits_{~~\Delta s} n_{\phi}^{(0)}ds,
\end{equation}\mbox{}
are the total signal photon flux and the total noise photon flux passing
through the receiving surface $\Delta s$, respectively.\\
\indent It should be pointed out that, as one very important feature of the BPF expressed by the Gaussian-type wave beam solutions, Eqs. (\ref{eqa2}) to (\ref{eqa7}),   there is no BPF $n^{(0)}_x$ propagating along the $x$-direction (i.e., a direction perpendicular to the symmetrical axis $z$ of the GB) due to $\tilde E^{(0)}_y=\psi_{ey}=0$ [see Eq. (\ref{eqa2})], namely, 
\begin{eqnarray}
	\label{eqb3}
	n_x^{(0)}=\frac{1}{2\mu_0\hbar\omega_e}Re\langle{\tilde{E}}_y^{(0)*}{\tilde{B}}_{z}^{(0)}\rangle=\frac{1}{2\mu_0\hbar\omega_e}Re\langle\tilde\psi_{ey}^*\psi_{bz}\rangle=0.~~~~~~~~
\end{eqnarray}
\indent However, the PPFs $n_x^{(1)}$ propagating along the $x$-direction has significant non-vanishing values. According to Eqs. (\ref{eq29}) and  (\ref{eq63}), we have
\begin{eqnarray}
	\label{eqb4}
	&& n_{x-\otimes}^{(1)}=\frac{1}{2\mu_0\hbar\omega_e}Re\langle{\tilde{E}}_y^{(1)*}{\tilde{B}}_{z}^{(0)}\rangle_{\omega_e=\omega_g}=\frac{1}{\mu_0\hbar\omega_e}\nonumber\\
	&&\cdot\{
	\frac{A_{\otimes}\hat{B}_y^{(0)}\psi_0k_g\Delta \textit{z}~r}{2[1+(z/f)^2]^{\frac{1}{2}}(z+f^2/z)}
	\sin[\frac{k_e r^2}{2R} -\tan ^{-1}(\frac{z}{f})+\delta]\nonumber\\
	&&+\frac{A_{\otimes}\hat{B}_y^{(0)}
		\psi_0\Delta\textit{z}~r}{W_0^2[1+(z/f)^2]^{\frac{3}{2}}} 
	\cos[\frac{k_e r^2}{2R} -\tan ^{-1}(\frac{z}{f})+\delta]\}exp(-\frac{r^2}{W^2})\sin\phi,~~~~~~~~~~
\end{eqnarray}
for the coupling between the transverse static magnetic field $\hat{B}_y^{(0)}$ and the GB, and we have
\begin{eqnarray}
	\label{eqb5}
	&&	n_{x-x}^{(1)}=\frac{1}{\mu_0\hbar\omega_e}Re\langle{\tilde{E}}_y^{(1)*}{\tilde{B}}_{z}^{(0)}\rangle_{\omega_e=\omega_g}=\frac{1}{\mu_0\hbar\omega_e}\nonumber\\
	&&\cdot \{
	\frac{A_{x}\hat{B}_z^{(0)}\psi_0k_g\Delta \textit{z}~r}{2[1+(z/f)^2]^{\frac{1}{2}}(z+f^2/z)}
	\sin[\frac{k_e r^2}{2R} -\tan ^{-1}(\frac{z}{f})+\delta]\nonumber
\end{eqnarray}
\begin{eqnarray}
	&&+\frac{A_{x}\hat{B}_z^{(0)}
		\psi_0\Delta\textit{z}~r}{W_0^2[1+(z/f)^2]^{\frac{3}{2}}}\cos[\frac{k_e r^2}{2R} -\tan ^{-1}(\frac{z}{f})+\delta]\} \nonumber\\
	&&\cdot\exp(-\frac{r^2}{W^2})\sin\phi,
\end{eqnarray}
for the coupling between the longitudinal static magnetic field $\hat{B}_z^{(0)}$ and the GB.\\ 
\indent Eqs. (\ref{eqb4}) and (\ref{eqb5}) show that the signal photon fluxes received in this direction would be the PPFs generated by the pure $\otimes$-type polarization (the tensor-mode gravitons) and by the pure  $x$-type polarization (the vector-mode gravitons), respectively.\\
\indent Since the  $n_x^{(0)}=0$, the noise photon flux fluctuation in the displaying condition, Eq. (\ref{eqb1}), will not be caused by the BPF of this direction, but would be  caused by other noise photon fluxes. However, the latter are much less than the former\cite{jmp498} in the 3DSR. In this case, the displaying condition in Table \ref{tbb1}  can be greatly relaxed. In other words, the PPFs produced by the HFGWs in the braneworld\cite{cqgF33} and in the short-term anisotropic inflation\cite{Ito.anisotropic.2016}, can almost be instantaneously  displayed, and the requisite minimal accumulation time of the signal displaying the HFGWs predicted by the interaction\cite{prd044017} of astrophysical plasma with intense EM waves, by the pre-big-bang\cite{pr373,sciam290} and by the quintessential inflationary\cite{prd123511, cqg045004,Giovannini2014} models, would be further relaxed effectively. Thus, utilizing a highly orientational receiving surface to display the PPFs, will be very useful.\\
\indent By the way, the wave impedance and wave impedance matching would also play important roles for the displaying condition. For most of the signal photon fluxes discussed in this paper, ratios of the electric components to the magnetic components are much less than that of the background noise photon fluxes. The typical values of the wave impedance of the former are $\sim10^{-4}\Omega$ or less, and the typical values of the latter are $\sim100\Omega$ to $377\Omega$\cite{LiNPB2016,Haslett}. This means that the 3DSR system would be equivalent to a ``good superconductor'' to the PPFs. Thus, the PPFs could be distinguished from the BPF and other noise photons by the wave impedance matching, and this is another important symbol to distinguish them.\\
\indent It is very interesting to compare the PPFs $n_{x-\otimes}^{(1)}$, $n_{x-x}^{(1)}$ propagating along the x-direction and the PPF $n_{y-\oplus,b,l}^{(1)}$ propagating along the y-direction. Notice that in the x-direction the $n_{x}^{(0)}=0$ (the dominant noise photon flux in the x-direction caused by the GB) [see Eq. (\ref{eqb3})], while the $n_{y}^{(0)}$ (the dominant noise photon flux in the y-direction caused by the GB) has non-vanishing value (see below), thus the displaying condition of the PPFs $n_{x-\otimes}^{(1)}$, $n_{x-x}^{(1)}$, Eqs. (\ref{eqb4}), Eqs. (\ref{eqb5}),	  is much better than that of $n_{y-\oplus,b,l}^{(1)}$, but distinguishing and displaying the $n_{y-\oplus,b,l}^{(1)}$ are still possible due to very different physical behavior between the $n_{y-\oplus,b,l}^{(1)}$ and $n_{y}^{(0)}$ (see below).\\
\indent From  Eq. (\ref{eq90}) and Eqs. (\ref{eqa2}), (\ref{eqa3}), we have 
\begin{eqnarray}
	\label{eqb6}
	&& n_{y-\oplus,b,l}^{(1)}=\frac{1}{2\mu_0\hbar\omega_e}Re\langle{\tilde{E}}_z^{(0)*}{\tilde{B}}_{x}^{(1)}\rangle_{\omega_e=\omega_g}=\hat{B}_{x}^{(0)}k_g\Delta \textit{z}~x
	\nonumber\\
	&& \cdot\frac{(A_b-A_{\oplus}+2\sqrt{2}A_l)}{2\mu_0\hbar\omega_e}Re\langle F_1^*\cdot \exp[i(k_gz-\omega_gt)]  \rangle_{\omega_e=\omega_g}, ~~~~~~~
\end{eqnarray}
\begin{eqnarray}
	\label{eqb7}
	\text{and,}~~~n_{y}^{(0)}&=&\frac{1}{2\mu_0\hbar\omega_e}Re\langle{\tilde{E}}_x^{(0)*}{\tilde{B}}_{z}^{(0)}\rangle_{\omega_e=\omega_g}
	\nonumber\\
	&=&\frac{\psi_0^2y}{2\mu_0\hbar\omega_e}exp(-\frac{2r^2}{W^2})[\frac{k_e}{\omega_e\sqrt{1+(z/f)^2}}],~~~
\end{eqnarray}
Eqs.  (\ref{eqb6}) and  (\ref{eqb7}) show following important properties:
\begin{eqnarray}
	\label{eqb8}
	\text{(i).}~n_{y-\oplus,b,l}^{(1)}\mid_{y=0}=n_{y-\oplus,b,l(max)}^{(1)},~
	\text{while,}~n_{y}^{(0)}\mid_{y=0}=0, ~~~~~~
\end{eqnarray}
i.e., the position (the longitudinal symmetrical surface of the GB) of peak value of the signal photon flux $n_{y-\oplus,b,l}^{(1)}$ is just the zero value area of the BPF $n_{y}^{(0)}$ (the background noise photon flux).
\begin{eqnarray}
	\label{eqb9}
	\text{(ii).}~~~~~n_{y-\oplus,b,l}^{(1)}\propto exp(-\frac{r^2}{W^2})= exp(-\frac{x^2+y^2}{W^2}),~~~~~
\end{eqnarray}
\begin{eqnarray}
	\label{eqb10}
	n_{y}^{(0)}\propto exp(-\frac{2r^2}{W^2})= exp(-\frac{2(x^2+y^2)}{W^2}),
\end{eqnarray}
i.e., the decay rate of $n_{y-\oplus,b,l}^{(1)}$ is obviously slower than that of $n_{y}^{(0)}$, although the peak value of $n_{y}^{(0)}$ is much larger than that of  $n_{y-\oplus,b,l}^{(1)}$. \\
(iii). $n_{y-\oplus,b,l}^{(1)}$ [Eq. (\ref{eqb6})], is an even function of the coordinates $y$; thus, $n_{y-\oplus,b,l}^{(1)}$ has the same propagating direction in the regions of $y>0$ and $y<0$. At the same time, $n_{y}^{(0)}$ (Eq. (\ref{eqb7})), is an odd function of the coordinates $y$, so the propagating directions of  $n_{y}^{(0)}$ are opposite in the regions of $y>0$ and $y<0$ [see Fig. (\ref{figureb1})]. These properties provide the distinguishability between $n_{y-\oplus,b,l}^{(1)}$ and $n_{y}^{(0)}$ [see Fig. (\ref{figureb1})].\\
\indent Fig.  \ref{figureb1} shows that although the peak of  $n_{y}^{(0)}$ is much larger than that of $n_{y-\oplus,b,l}^{(1)}$ (the signal photon flux generated by the HFGWs in the braneworld\cite{cqgF33}), it is always possible to find suitable positions, where $n_{y-\oplus,b,l}^{(1)}$ and $n_{y}^{(0)}$ are comparable and distinguishable, and  $n_{y-\oplus,b,l}^{(1)}$ would be measurable. In this case, due to the background photon flux decays faster than the signal photon flux, in the area around $y=0.25m$, the $n_{y}^{(0)}$ already decays into the comparable level to the  $n_{y-\oplus,b,l}^{(1)}$, and in the further area, the background noise will decay into lower level than the measurable signal photon flux. This is satisfactory.\\

\section{The matching function between the HFGWs and the 3DSR system.}
\label{appendixC}
\indent Here we choose $n_{x-\otimes}^{(1)}$, Eq. (\ref{eqb4}), as an example to calculate the total signal photon flux propagating along the $x$-direction at effective receiving surface $\Delta s$. Then the  $n_{x-\otimes(total)}^{(1)}$  can also be expressed as 
\begin{eqnarray}
	\label{eqc1}
	|n_{x-\otimes(total)}^{(1)}|=\frac{A_{\bigotimes}\hat{B}_y^{(0)}\psi_0}{\mu_0\hbar\omega_e}|f(k_g, \bf{x})|,
\end{eqnarray}
where
\begin{eqnarray}
	\label{eqc2}
	&&	f(k_g, \textbf{x})=\int_{\Delta s} \{	\frac{k_g \Delta z r}{2[1+(z/f)^2]^{\frac{1}{2}}(z+f^2/z)}\nonumber\\
	&&	\cdot\sin[\frac{k_e r^2}{2R} -\tan ^{-1}(\frac{z}{f})+\delta]+\frac{\Delta zr}{W_0^2[1+(z/f)^2]^{\frac{3}{2}}}\nonumber\\
	&&	 \cdot \cos[\frac{k_e r^2}{2R} -\tan ^{-1}(\frac{z}{f})+\delta]\}\exp(-\frac{r^2}{W^2})\sin\phi dydz, ~~~~~
\end{eqnarray}	
is the matching function in Cartesian coordinate system, where $r=\sqrt{x^2+y^2}$, $\sin\phi=y/r=y/\sqrt{x^2+y^2}$. From  Eq. (\ref{eqc1}), we have
\begin{eqnarray}
	\label{eqc3}
	|n_{x-\otimes(total)}^{(1)}|_{max}=\frac{A_{\bigotimes}\hat{B}_y^{(0)}\psi_0}{\mu_0\hbar\omega_e}|f(k_g, \bf{x})|_{max},
\end{eqnarray}
\begin{figure}[h]
	\centerline{\includegraphics[scale=0.5]{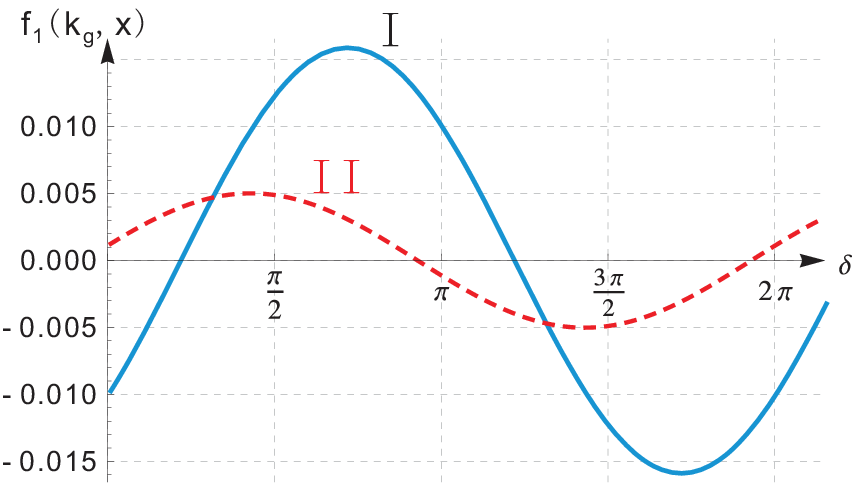}} 
	\caption{\textbf{Relation curves between the matching function $f(k_g, x)$ and the phase $\delta$.} In fact, $\delta$ is the phase difference between the non-stochastic HFGW and the GB in the 3DSR system. It is shown that for the parameters: $\nu_e=\nu_g=3\times10^9 Hz$, $W_0=0.06m$, $y\in[0m,~0.1m]$, $z\in[0,~0.3m]$, (i.e., $\Delta s=3.0\times 10^{-2}m^2$) and the matching function $f(k_g, x)$  at the longitudinal symmetrical surface (i.e., the surface of $x=0$), then $\delta=18\pi/25$ (or 2.26195) is the optimal phase, and $f(k_g, {\bf{x}})=0.0159$ is the best matching function. For the matching function $f(k_g, x)$  at the surface of $x=10 cm$), $\delta=43\pi/100$ (or 1.35088) is the optimal phase, then $f_1(k_g, x)=0.00502$ is the best matching function.}
	\label{figurec1}
\end{figure}
where $|n_{x-\otimes(total)}^{(1)}|_{max}$ corresponds to the best coupling state. For the typical parameters of the 3DSR system, $A_{\otimes}=10^{-23}$, $\nu_g=3\times10^9Hz$ (the amplitude and frequency of HFGWs in the braneworld\cite{cqgF33});   $\hat{B}_y^{(0)}=10 T$, $\psi_0=1.95\times10^{3}Vm^{-1}$ (for the GB with $p=10 W$, $W_0=0.06m$, $y\in[0,~0.1m]$, $z\in[0,~0.3m]$, the relation curve between the matching function $f_1(k_g, x)$ and the phase $\delta$ can be shown in Fig.  \ref{figurec1}  (at the surfaces of $x=0$ and at the surface of $x=10cm$), i.e., the curves I and II respectively.\\
\indent For such non-stochastic constant amplitude HFGW, the optimal phase is $\delta=18\pi/25$ (or $2.26195$), then  $|f(k_g,{\bf{x}})|=|f(k_g,{\bf{x}})|_{max}=0.0159$, at the longitudinal symmetrical surface of $x=0$, and $|n_{x-\otimes(total)}^{(1)}|_{max}\approx1.21\times10^9s^{-1}$ at  receiving surface $\Delta s\sim3\times10^{-2}m^2$.\\
\indent For the stochastic HFGWs (e.g., the relic HFGWs in the pre-big-bang\cite{pr373,sciam290} or in the quintessential inflationary\cite{prd123511, cqg045004,Giovannini2014}), it is necessary to integrate over all possible phases. Notice that this integration is not equal to zero and it is only limited in the positive value region or in the negative value region (which correspond to two opposite propagating directions of the $n^{(1)}_{x-\otimes}$) due to the high directivity of the signal photon flux $n^{(1)}_{x-\otimes}$ and suitable orientation of the receiving surface. For the relic HFGW ($\nu_g=3\times10^9Hz$, $h\sim10^{-29}$) in the pre-big-bang, $|n_{x-\otimes(total)}^{(1)}|_{max}\approx 10^3~s^{-1} $, and for the relic HFGW ($\nu_g=3\times10^9Hz$, $h\sim10^{-30}$) in the quintessential inflationary,  $|n_{x-\otimes(total)}^{(1)}|_{max}\approx 10^2~s^{-1} $ (see Table \ref{tbb1}).\\
\indent In this case, the zero value characteristics of the background noise photo flux BPFs  $n^{(0)}_x$,  $n^{(0)}_y$ at the longitudinal symmetric surface, and the peak property of the signal photon fluxes ( $n^{(1)}_{x-\otimes}$,  $n^{(2)}_{x-\otimes}$) at the longitudinal symmetric surface,  inspire us to utilize the fractal membranes (they are very effective microwave lenses with strong focusing function and ``one-way valve'' property\cite{Houbo2015,FYLi_EPJC_2008,26,27,28} to the photon fluxes in the GHz to THz band), and they will provide better selecting way for the signal photon fluxes, especially for the signal photon fluxes generated by the stochastic relic HFGWs. These issues will also be discussed and studies in detail elsewhere.\\

\begin{acknowledgments}
This project is supported by the National Natural Science Foundation of China (Grant No.11375279, No.11605015 and No.11873001), the Foundation of China Academy of Engineering Physics No.2008 T0401 and T0402, the Science and Technology Research Program of Chongqing Municipal Education Commission (Grant No. KJQN201800105), and Natural Science Foundation Project of Chongqing cstc2018jcyjAX0767.  We very thank helpful discussions with Prof. Robert M. L. Baker, Prof. R. C. Woods, Prof. Guang-Li Kuang, Prof. Lian-Fu Wei,Prof. C.Corda, Tan, Prof. Yun-Fei Tan, Dr. Gary V. Stephenson, Dr. Andrew Beckwith and Dr. Eric W. Davis.
\end{acknowledgments}

\bibliographystyle{apsrev4-1}
\bibliography{arXivV5}

\end{document}